\documentclass[aps,preprint,floatfix,nofootinbib,showpacs]{revtex4-1}
\pdfoutput=1
\usepackage{graphicx,color}
\usepackage{hyperref}
\usepackage{amsmath}
\usepackage{amsfonts}
\usepackage{amssymb}
\usepackage{setspace}
\usepackage{feynarts}

\newenvironment{Eqnarray}
         {\arraycolsep 0.14em\begin{eqnarray}}{\end{eqnarray}}

\def\bit{\begin{itemize}}
\def\eit{\end{itemize}}
\def\ben{\begin{enumerate}}
\def\een{\end{enumerate}}
\def\beq{\begin{equation}}
\def\eeq{\end{equation}}

\def\bea{\begin{Eqnarray}}
\def\eea{\end{Eqnarray}}

\def\beq{\begin{equation}}
\def\eeq{\end{equation}}
\def\beqa{\begin{Eqnarray}}
\def\eeqa{\end{Eqnarray}}

\def\beq{\begin{equation}}
\def\eeq{\end{equation}}
\def\ifmath#1{\relax\ifmmode #1\else $#1$\fi}

\def\ls#1{\ifmath{_{\lower1.5pt\hbox{$\scriptstyle #1$}}}}
\def\lss#1{\ifmath{^{\,\lower2.5pt\hbox{$\scriptstyle #1$}}}}

\def\lsim{\mathrel{\raise.3ex\hbox{$<$\kern-.75em\lower1ex\hbox{$\sim$}}}}
\def\gsim{\mathrel{\raise.3ex\hbox{$>$\kern-.75em\lower1ex\hbox{$\sim$}}}}
\def\ifmath#1{\relax\ifmmode #1\else $#1$\fi}

\begin{document}
\begin{center}

\vspace*{1.5cm}
{\Large\bf Radiative corrections to the 
Triple Higgs Coupling in the Inert Higgs Doublet Model}

\vspace*{0.8cm}

\renewcommand{\thefootnote}{\fnsymbol{footnote}}

{\large 
Abdesslam Arhrib$^{1,2}$\footnote[1]{Email: aarhrib@ictp.it},
Rachid Benbrik$^{3,4}$\footnote[2]{Email: rbenbrik@ictp.it},
Jaouad El Falaki$^{1}$\footnote[3]{Email: jaouad.elfalaki@gmail.com},
Adil Jueid$^{1}$\footnote[4]{Email: ajueid@ictp.it}
}

\renewcommand{\thefootnote}{\arabic{footnote}}

\vspace*{0.8cm}
{\normalsize \it
$^1\,$D\'{e}partement de Math\'{e}matiques, 
Facult\'{e} des Sciences et Techniques,
Universit\'{e} Abdelmalek Essaadi, B. 416, Tangier, Morocco.\\
$^2\,$Physics Division, National Center for Theoretical Sciences,
  Hsinchu 300, Taiwan.\\
$^3\,$Faculty of Polydisciplinaire de Safi, Sidi Bouzid B.P 4162, 
46000 Safi, Morocco.\\
$^4\,$LPHEA, Faculty of Science Semlalia, Cadi Ayyad University, 
Marrakesh, Morocco.
}
\end{center}
\begin{abstract}
We investigate the implication of the recent discovery of a Higgs-like 
particle in the first phase of the LHC Run 1 on the 
Inert Higgs Doublet Model (IHDM). 
The determination of the Higgs couplings to SM particles 
and its intrinsic properties will get improved during 
the new LHC Run 2 starting this year. The new LHC Run 2 would also 
shade some light on the triple Higgs coupling. Such measurement is very
important in order to establish  the details of the electroweak 
symmetry breaking mechanism. 
Given the importance of the Higgs couplings both at the LHC and $e^+e^-$ 
 Linear Collider machines, accurate theoretical predictions are required.
We study the radiative corrections to the triple Higgs coupling $hhh$ 
and to $hZZ$, $hWW$ couplings in the context of the IHDM. 
By combining  several theoretical and experimental constraints on parameter
space, we show that extra particles might modify the triple Higgs coupling near
threshold regions. 
Finally, we discuss the effect of these corrections on the
double Higgs production signal at the $e^+e^-$ LC and show that they can 
be rather important.
\end{abstract}
\maketitle

\section{Introduction}\label{sec:intro}
The discovery of a new particle with a mass around 125-126 GeV in the search
for the Standard Model (SM) Higgs boson~\cite{Englert:1964et} was announced
simultaneously by the ATLAS and CMS collaborations in July
2012~\cite{atlas1,atlas2,atlas3,cms1,cms2,cms3}. Since then, 
more data has been taken and analyzed at the LHC. 
One of the primary goals of the Higgs groups at the LHC is now to 
study the properties of this new
resonance and determine if it is indeed the state predicted by the SM.

With this new discovery, a program of precision measurement involving the
Higgs boson has just started and will get improved with the new run of LHC and
future $e^+e^-$ Linear Collider (LC).
In fact, the $7\oplus 8$ TeV data allow the measurement of the Higgs 
couplings to gauge bosons and $\tau^+\tau^-$ with about 
20-30\% of precision while the Higgs couplings to $b\bar{b}$ and $t\bar{t}$ 
still suffer large uncertainties of  about 40-50\%. All these measurements 
will be improved with the new run of LHC at 13-14 TeV and the future
 $e^+e^-$ LC. \\

In order to confirm that the discovered Higgs-like particle is the 
SM Higgs responsible for the electroweak symmetry breaking, we need to 
know all its couplings to SM particles with 
accurate precision and also measure the trilinear and quartic 
self-couplings of the Higgs in order to be able to reconstruct 
the scalar potential.  
In this regards, the LHC with its high luminosity option have also the 
capability of measuring the SM triple Higgs couplings through one of the
following channels $gg\to hh\to
b\bar{b}\gamma\gamma, b\bar{b}\tau^+ \tau^-, 
b\bar{b} W^\pm W^{\mp *}$ \cite{double1,double2}. 
The $e^+e^-$ LC, which will provide some 
high precision measurement of the Higgs mass and its properties such as
couplings to SM particles and quantum numbers, 
 would also be able to perform SM triple Higgs coupling through 
$e^+e^-\to Zhh$ (double Higgs-Strahlung) and $e^+ e^-\to \nu_e\bar{\nu}_e hh$
 (WW fusion) with more than 700 GeV center of mass energy 
with better precision \cite{lc}. 
In the double Higgs-Strahlung process $Zhh$ the Z boson will
be reconstructed from $l^+l^-$ or $q\bar{q}$ pairs, while for the WW fusion
process the two Higgs can be reconstructed from $b\bar{b}b\bar{b}$ or 
$b\bar{b}$  and $W^+W^-$. \\

The discovery of this Higgs-like particle resonance opens a new era in  
elementary particle physics and leads to several theoretical and  
phenomenological studies on Higgs physics both in the SM and beyond. 
One of the very simplest extension of SM is the IHDM proposed, 
more than three decade ago, by Deshpande and
Ma \cite{Deshpande:1977rw} for electroweak symmetry breaking purpose.
Recently, the IHDM model has been very attractive because it provides a
 dark matter candidate \cite{idm_dm,idm_dm2}, generates tiny neutrino 
masses \cite{Ma:2006km} and also solve the naturalness 
problem \cite{Barbieri:2006dq}. 
Phenomenology of IHDM have been  extensively studied during last 
decade \cite{idm_dm2,artc,stal}.\\

The aim of this paper is to study the effect of the one loop 
radiative corrections to the triple Higgs coupling $hhh$ 
as well as $hZZ$ coupling 
in the framework of the IHDM. We will also compute the well 
known SM radiative corrections 
to the triple Higgs coupling as check of our procedure.
Once these effect are well studied, we then proceed to the evaluation of 
radiative corrections to the double Higgs Strahlung process $e^+e^- \to Zhh$. 
For this purpose, we will apply an on-shell 
renormalization scheme to evaluate these one-loop corrections. For the
numerical evaluation, we will take into account all the theoretical and
experimental constraints on the scalar sector of the Model.\\

The paper is organized as follow: in the second section we introduce the 
IHDM model and describe the theoretical and experimental constraints 
that the model is subject to. In the third section we introduce the on-shell 
renormalization scheme for the triple Higgs coupling $hhh$ and $hZZ$ 
in the IHDM and present our numerical results in the fourth section. 
Numerical analysis of the double
Higgs-strahlung is presented in the fifth section. Our conclusion is given
in the last section.


\section{The Inert Higgs Doublet Model}
\subsection{The Model}
The IHDM is one of the most simplest models for the scalar dark matter, a
version of a two Higgs double model with an exact $Z_2$ symmetry. 
The SM scalar sector is extended by an inert scalar doublet 
 $H_2$ which can provide a stable dark
matter candidate. Under $Z_2$ symmetry all the  SM particles are even while
$H_2$ is odd and it could mix with the SM-like Higgs doublet.
We shall use the following parameterization of the two doublets :
\begin{eqnarray}
H_1 = \left (\begin{array}{c}
G^\pm \\
\frac{1}{\sqrt{2}}(v + h + i G^0) \\
\end{array} \right)
\qquad , \qquad
 H_2 = \left( \begin{array}{c}
H^\pm\\ 
\frac{1}{\sqrt{2}}(H^0 + i A^0) \\ 
\end{array} \right) 
\end{eqnarray}
with $G^0$ and $G^\pm$ are the Nambu-Goldstone bosons absorbed by the
longitudinal component of $W^\pm$ and $Z^0$,respectively. $v$ is the
vacuum expectation value (VEV) of the SM Higgs $H_1$. Within the IHDM the
scalar doublet $H_2$ does not couple with the SM fermions and therefore the
$H_2$-fermions interaction are present  only through mixing with $H_1$.
The most general renormalizable, gauge invariant and 
CP invariant potential is given by :
\begin{eqnarray}
V  &=&  \mu_1^2 |H_1|^2 + \mu_2^2 |H_2|^2  + \lambda_1 |H_1|^4
+ \lambda_2 |H_2|^4 +  \lambda_3 |H_1|^2 |H_2|^2 + \lambda_4
|H_1^\dagger H_2|^2 \nonumber \\
&+&\frac{\lambda_5}{2} \left\{ (H_1^\dagger H_2)^2 + {\rm h.c} \right\} 
\label{potential}
\end{eqnarray}
In the above potential there is no mixing terms like $\mu_{12}^2 (H_1^\dagger H_2 +
h.c)$ because of the unbroken $Z_2$ symmetry. By hermicity of the potential,
 all $\lambda_i, i = 1, \cdots, 4$  parameters are real. The phase of 
$\lambda_5$ can be absorbed by a suitable redefinition 
 of the fields $H_1$ and $H_2$, therefore the scalar sector is CP conserving.
 After spontaneous symmetry breaking of $SU(2)_L \otimes U(1)_Y$ 
 down to $U(1)_Q$, the spectrum of 
this potential will have five scalar particles: two CP even $H^0$ and 
$h$ which will be identified as the SM Higgs boson, a CP odd $A^0$ and  
a pair of charged scalars $H^\pm$. 
Their masses are given by:
\begin{eqnarray}
&& m_h^2 = - 2 \mu_1^2 = 2 \lambda_1 v^2 \nonumber \\
&& m_{H^0}^2 = \mu_2^2 + \lambda_L v^2 \nonumber \\
&&  m_{A^0}^2 = \mu_2^2 + \lambda_S v^2 \nonumber \\
&&  m_{H^{\pm}}^2 = \mu_2^2 + \frac{1}{2} \lambda_3 v^2
  \label{spect.IHDM}
\end{eqnarray}
where $\lambda_{L,S}$ are defined as:
\begin{eqnarray}
\lambda_{L,S} &=& \frac{1}{2} (\lambda_3 + \lambda_4 \pm \lambda_5)
\end{eqnarray}
This model involves $8$ independent parameters: five $\lambda$, two $\mu_i$ 
and $v$. One parameter is eliminated by the minimization condition 
and the VEV is fixed by the $W$ boson mass. Finally, we are left with six 
independent  parameters which we choose as follow :
\begin{eqnarray}
 \{ \mu_2^2, \lambda_2, m_h, m_{H^\pm}, m_{H^0}, m_{A^0} \}
 \label{param.IHDM}
\end{eqnarray}

\subsection{Theoretical and Experimental Constraints}
In order to have vacuum stability, the parameters of the potential need to
satisfy the positivity conditions. Namely, the potential should be bounded
from below in all the directions of the field space, i.e. should not go to
negative infinity for large field values. We have this set of constraints :
\begin{eqnarray}
 \lambda_1>0 \quad , \quad \lambda_2>0 \quad , \quad  \lambda_3 + 2
 \sqrt{\lambda_1 \lambda_2} > 0 \quad \quad and \quad \quad
 \lambda_3 + \lambda_4 - |\lambda_5| > 2 \sqrt{\lambda_1 \lambda_2} 
\end{eqnarray}
We ask that the perturbative unitarity is maintained in variety of 
scattering processes at high energy: 
scalar-scalar, scalar-vector, vector-vector. 
Using the equivalence theorem which replaces the $W$ and $Z$ bosons by 
the Goldstone bosons. We find a set of four matrices with entries are 
the quartic couplings, the diagonalization of these
matrices gives us this set of eigenvalues \cite{Unitarity}:
\begin{eqnarray}
 e_{1,2} = \lambda_3 \pm \lambda_{4} \textrm{ , } e_{3,4} = \lambda_3 \pm \lambda_{5} \nonumber \\
 e_{5,6} = \lambda_3  + 2 \lambda_4 \pm 3 \lambda_5 \nonumber \\
 e_{7,8} = - \lambda_1 - \lambda_2 \pm \sqrt{(\lambda_1 + \lambda_2)^2 + \lambda_4^2} \nonumber \\
 e_{9,10} = -3 \lambda_1 - 3 \lambda_2 \pm \sqrt{9(\lambda_1 - \lambda_2)^2 + (2 \lambda_3 + \lambda_4)^2}  \nonumber \\
 e_{11,12} = - \lambda_1 - \lambda_2 \pm \sqrt{(\lambda_1 - \lambda_2)^2 + \lambda_5^2}
\end{eqnarray}
By requiring that $e_i \leq 8 \pi$, we find the strongest constraint on
$\lambda_{1,2}$ to be: 
$\lambda_{1,2} \leq \frac{4 \pi}{3}$. We also require that the parameters of 
the scalar potential remains perturbative, i.e: 
$|\lambda_i| \leq 8 \pi$.

In order to have an inert vacuum the following constraint 
 should be satisfied \cite{maria}:
\begin{eqnarray}
 m_h^2, m_{H^\pm}^2, m_{H^0}^2, m_{A^0}^2 > 0 \quad \quad \textrm{ and } 
\quad \quad v^2 > -\frac{\mu_2^2}{\sqrt{\lambda_1 \lambda_2}}
\end{eqnarray}
The extra scalar particles affects quantum corrections to the 
$W$ and $Z$ bosons self energies. The corrections are parameterized by 
 the oblique parameters,
$S$, $T$ and $U$~\cite{Peskin} which are constrained from electroweak
precision measurements. Taking the reference Higgs mass as $m_h = 125 $ GeV 
and $m_t = 173$ GeV, the tolerated ranges are found at 
 fixed U=0 \cite{Baak:2014ora}:
\begin{eqnarray}
 \Delta S = 0.06\pm0.09, \quad\quad\quad {and}\quad\quad\quad 
 \Delta T = 0.10\pm0.07
\end{eqnarray}
with correlated factor of +0.91.
Where $\Delta S = S^{\textrm{IHDM}} - S^{\textrm{SM}}$ and 
$\Delta T = T^{\textrm{IHDM}} - T^{\textrm{SM}}$. 
The formulas for $\Delta S$ and $\Delta T$ in the IHDM can be found in 
Ref~{\cite{htogaga,Barbieri:2006dq}}. 

Searches of scalar particles~\footnote{these dark Higges
have the same signature like neutralinos and charginos of the Minimal
Supersymetric Standard Model (MSSM).} of the IHDM at colliders 
\cite{Belanger:2015kga} is not directly 
performed yet. However, several studies 
 \cite{Dolle:2009ft,Miao:2010rg,Gustafsson:2012aj}
 applied SUSY searches involving two, three or multiple 
leptons with missing transverse energy $E^{miss}_T$ to the case of IHDM
 and set some limits on the dark Higges. 
We choose in our analysis~\footnote{if $H$ or $A$ is DM candidate then its mass could be as low as few GeV.} :
\begin{eqnarray}
 m_\Phi \geq 100 \textrm{ GeV } \quad \textrm{ where } \Phi = H^\pm, H^0, A^0
\end{eqnarray}
Finally, the magnitude of a possible Higgs boson signal at the LHC is
characterized by the signal strength modifier, defined as $R_{\gamma\gamma}$ by :
\begin{eqnarray}
 R_{\gamma\gamma} = \frac{\sigma(pp \to h \to 
\gamma\gamma)^{\textrm{IHDM}}}{\sigma(pp \to h_{\textrm{SM}} \to \gamma\gamma)} 
 = \frac{Br(h \to \gamma \gamma)^{\textrm{IHDM}}}{Br(h_{\textrm{SM}} 
\to \gamma \gamma)}
\end{eqnarray}
$h_{\textrm{SM}}$ denotes a 125 GeV SM Higgs boson. In our analysis below,
while we will show  points which satisfy theoretical and experimental
constraints from our scans, we will highlight the
points for which $R_{\gamma\gamma}$ is consistent with the measured
$\mu_{\gamma\gamma}$ at the LHC. The latest publicly available measurements
read \cite{cmsreport,Aad:2014eha}
\begin{eqnarray}
 \mu_{\gamma \gamma}^{\rm{CMS}} &=&1.13\pm 0.24 \\
 \mu_{\gamma \gamma}^{\rm{ATLAS}}&=&1.17 \pm 0.27 
\end{eqnarray}

\section{Radiative corrections to triple Higgs couplings $hhh$ and $hZZ$}
In this section we calculate the one-loop radiative corrections 
to the trilinear Higgs coupling $hhh$ and $hZZ$ in the SM and IHDM. 
The correction to those two couplings are part of the correction to double
Higgs strahlung process $e^+e^- \to Zhh$ which we will discuss in 
section V. Those couplings are given at the tree-level by:
\begin{eqnarray}
&& \lambda_{hhh} = \frac{-3 m_h^2}{v} \label{tree}\\
&& hZ_\mu Z_\nu = \frac{e m_W}{s_W c_W^2}g_{\mu\nu}\label{tree-hZZ}
\end{eqnarray}
As one can see, both couplings $hhh$ and $hZZ$ involve 
only SM parameters.
Those couplings receive corrections from one-loop diagrams.
The one-loop effects from the SM particles have been studied in 
\cite{kanemura,kanemura1,penaranda} for $hhh$ and in \cite{Kniehl:1990mq} 
for $hZZ$. These effects are dominated by the top quark loops
which does not exceed $10 \%$ for $hhh$ and $1.5\%$ for $hZZ$. 

New physics effects to $hhh$ coupling have been analyzed in the context of the 
Two Higgs Doublet Model \cite{kanemura} and the MSSM \cite{penaranda}. 
It was found that these effects can enhance significantly this 
coupling in a wide range of parameter space. Furthermore, 
these corrections depend on the model and hence
any deviation from the SM tree level relation (\ref{tree}) 
by more than $10 \%$ would be an evidence for the presence 
of new physics.

The coupling $hZZ$ have been analyzed in the framework of the two Higgs
doublet model \cite{kanemura1} and it has been found that the effect is 
rather small $1$\% to $2$\%.

\begin{figure}[ptb]
\resizebox{167.5mm}{!}{\includegraphics{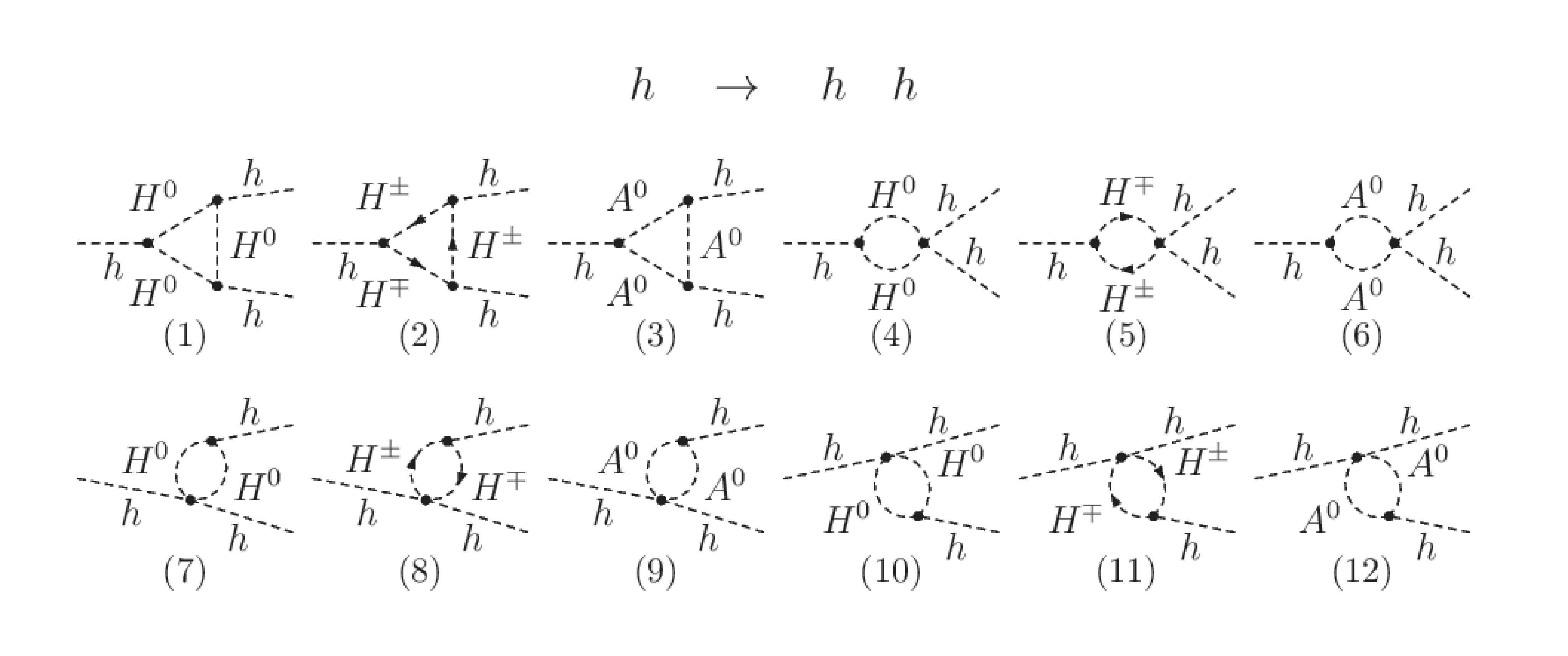}}
\caption{One loop Feynman diagrams contributing to  $hhh$ in the IHDM. 
S stands for $H^\pm, G^\pm, G^0$ and V = $W^\pm$. }
\label{FD}
\end{figure}

We have calculated the radiative corrections to the tree level 
triple Higgs coupling $hhh$ and $hZZ$ both in the SM and IHDM 
in the Feynman gauge including all the particles of the model in the loops. 
The Feynman diagrams from IHDM contributing to $\lambda_{hhh}$ 
coupling are shown in Fig.(~\ref{FD}).

The one-loop amplitude are calculated using dimensional regularization.
The calculation was done with the help of FeynArts 
 and FormCalc \cite{FA2} packages. Numerical evaluation of the one-loop 
scalar integrals have  been done with LoopTools \cite{FF}.
We have checked both numerically and analytically the UV 
finiteness of the amplitudes. \\
In order to do that, we have considered $hhh$ and $hZZ$ at one-loop level:
\begin{itemize}
\item[i)] for $hhh$, we considered the decay of an off-shell 
Higgs boson into a Higgs boson pairs $h^*(q)\to h(k_1)h(k_2)$ 
at the one-loop level.
Where $q$, $k_1$ and $k_2$ are the 4-momenta of the three 
particles satisfying on shell conditions $k_1^2 = k_2^2 = m_h^2$ for 
final state Higgs pairs and an off shell condition $q^2 \neq m_h^2$
 for the decaying Higgs.
\item[ii)] For $hZZ$, we follow ref~\cite{kanemura1} and write:
\begin{eqnarray}  
 M^{\mu\nu}_{hZZ}(q^2=m_h^2,k_1^2, k_2^2)= M_1^{hZZ} g^{\mu\nu} 
                 + M_2^{hZZ} \frac{k_1^\nu k_2^\mu}{m_Z^2}
                 + M_3^{hZZ} \,i \epsilon^{\mu\nu\rho\sigma} 
                           \frac{{k_1}_\rho^{} {k_2}_\sigma^{}}{m_Z^2}, 
\label{eq:hzz-form}
\end{eqnarray}
where $k_1$ and $k_2$ are the momenta of outgoing $Z$ bosons. We assume that 
the decaying Higgs and one of the Z boson  are on-shell  $q^2=m_h^2$,
$k_1^2=m_Z^2$ while the other Z boson is off-shell $k_2^2=(m_h-m_Z)^2$.
Using power counting arguments, it is expected that 
$M_1^{hZZ}$ receives the highest power contribution of the 
heavy fermions    masses. 
Therefore, in what follow we will take into account only the 
 $M_1^{hZZ}$ form-factor to $hZZ$ coupling.
\end{itemize}

Since we are dealing with a processes at the one-loop level, 
a systematic treatment of the UV divergences have to be considered. 
We will use the on-shell renormalization scheme in 
which the input parameters coincide with 
the physical masses and couplings \cite{onshell1}. 
In the on shell scheme, a redefinition of the fields and parameters is
performed. This redefinition cast the Lagrangian into a bare 
Lagrangian and counter-term. 
The counter-terms are calculated by specific renormalization conditions 
which allow us to cancel the UV divergences of the diagrams with loops. 
Furthermore, since we have three Higgs as external particles and 
there is no mixing between the SM doublet $H_1$ and the inert 
doublet $H_2$, we do not need to renormalize the particle 
 content of the scalar potential of the
 IHDM. The tree level coupling $hhh$ eq.~(\ref{tree}) depends only 
 on Higgs mass and the vev as in the SM, then the renormalization 
procedure will be the same as in the SM \cite{onshell1}. 
We redefine the SM fields and parameters as follow:
\begin{eqnarray}
&& m_V^2 \to m_V^2 + \delta m_V^2 \quad, \quad V = W^\pm, Z \nonumber \\
&& m_h^2 \to m_h^2 + \delta m_h^2 \nonumber\\
&& s_W \to s_W + \delta s_W \nonumber \\
&& e \to (1 + \delta Z_e) e \nonumber\\ 
&& t \to t + \delta t \nonumber \\
&& W^\mu \to Z_W^{1/2} W^\mu = \left(1 + \frac{1}{2} \delta Z_W\right) 
W^\mu\nonumber \\
&& Z^\mu \to \left(1 + \frac{1}{2} \delta Z_{ZZ}\right) Z^\mu + 
\frac{1}{2} \delta Z_{ZA} A^\mu \nonumber\\
&& A^\mu \to \left(1 + \frac{1}{2} \delta Z_{AA}\right) A^\mu + 
\frac{1}{2} \delta Z_{AZ} Z^\mu \nonumber\\
&& h \to Z_h^{1/2} h = \left(1 + \frac{1}{2} \delta Z_h \right) h 
\end{eqnarray}

where $s_W = \sin \theta_W$ is the Weinberg angle and 
$t = v (\mu_1^2 - \lambda_1 v^2)$ is the tadpole which 
is zero at tree level once the minimization condition is used  
but will receives again finite radiative corrections at the 
one-loop level. To ensure that the VEV is the 
same in all orders of perturbation theory, it is well known that one need to
  renormalize  the Higgs tadpole: i.e, all Higgs tadpole amplitudes $T$ are 
absorbed into the  counter-term $\delta t$. Thus, we put the first condition:
\begin{equation}
 \hat{T} =  \delta t + T = 0
\end{equation}
The mass counter-terms are fixed by the on shell conditions \cite{onshell1}:
\begin{eqnarray}
&& \textrm{Re} \hat{\Sigma}_{T}^{VV}(m_V^2) = 0 \quad, \quad V=W,Z \nonumber \\
&& \textrm{Re} \hat{\Sigma}_h(m_h^2) = 0 
\end{eqnarray}
The field renormalization constants are fixed by imposing 
that the residue of the two point Green functions to be equal to unity 
 and the mixing $\gamma$-Z vanish for $k^2=m_Z^2$.
While the electric charge renormalization
constant $\delta Z_e$ is treated like in quantum electrodynamics
and is fixed from the $e^+e^- \gamma$ vertex. The renormalized 
three point function $\hat{\Gamma}_{e^+e^- \gamma}^\mu$
 satisfy at the Thomson limit:
$$\hat{\Gamma}^\mu_{e^+e^- \gamma} (\not{p}_1 = 
\not{p}_2 =m, q^2 = 0) = e $$ 
Furthermore, the counter-term $\delta s_W$ can be obtained 
from the on-shell definition $s_W^2 = 1 - \frac{m_W^2}{m_Z^2}$ as a function
of $\delta m_W$ and $\delta m_Z$.

Inserting these redefinitions  into the Lagrangian,
we find the following counter term for $hhh$ and $hZZ$ \cite{onshell1}:
\begin{eqnarray}
 &&\delta {\mathcal{L}}_{hhh} = \frac{-3e^2}{2 s_W} \frac{m_h^2}{m_W} 
\left(\delta Z_e - \frac{\delta s_W}{s_W} + \frac{\delta m_h^2}{m_h^2} + 
\frac{e}{2 s_W} \frac{\delta t}{M_W m_h^2} 
-\frac{\delta m_W^2}{2 m_W^2} + \frac{3}{2} \delta Z_h \right) h^3 \label{cthhh} \\
&&\delta {\mathcal{L}}_{hZZ} =\delta M_1=
\frac{e m_W}{s_W c_W^2} 
    \left(\delta Z_e + \frac{2 s_W^2 - c_W^2}{c_W^2 s_W}\delta s_W + 
\frac{\delta m_W^2}{2m_W^2} + \frac{1}{2} \delta Z_H + \delta Z_Z\right)
hZ^\mu Z^\nu
\end{eqnarray}
By adding the un-renormalized amplitude for $hhh$ and $hZZ$ 
to the above corresponding counter-terms, one find 
the renormalized amplitudes
\begin{eqnarray}
&&\hat{\Gamma}_{hhh}(q^2, m_h^2, m_h^2)= {\Gamma}_{hhh}^{1-loop}(q^2, m_h^2, m_h^2)+
\delta {\mathcal{L}}_{hhh}\nonumber\\
&&\hat{\Gamma}_{hZZ}(m_h^2,m_Z^2,(m_h-m_Z)^2)= {M_1}_{hZZ}^{tree}+
{M_1}_{hZZ}^{1-loop}+\delta {\mathcal{L}}_{hZZ}\nonumber\\
\end{eqnarray}
which becomes UV finite. For our numerical illustrations, 
we define the following ratios:
\begin{eqnarray}
 &&\Delta \Gamma_{hhh} = \frac{\textrm{Re}(\tilde{\Gamma}_{hhh}(q^2))}{\lambda_{hhh}} \\
&&\Delta \Gamma_{hZZ} = \frac{\hat{\Gamma}_{hZZ}^{IDM} -
   \hat{\Gamma}_{hZZ}^{SM}}{{M_1}_{hZZ}^{tree}}=\frac{M_1^{IHDM} - 
M_1^{SM-tree} }{M_1^{SM-tree} }
\end{eqnarray}
Where $\hat{\Gamma}_{hhh}$ is the renormalized vertex. 

\section{Numerical results}
\subsection{SM case}
In our numerical analysis, the parameters are chosen as follow : 
\begin{eqnarray}
 m_t = 173.5 \textrm{ GeV } \quad, \quad m_W = 80.3996 \textrm{ GeV } 
\quad, \quad 
m_Z = 91.1875 \textrm{ GeV} \quad, \quad   
\alpha = \frac{1}{127.934} \nonumber
\end{eqnarray}
and the on-shell definition of the Weinberg angle: 
$\sin^2 \theta_W=1-\frac{m_W^2}{m_Z^2}$.

\begin{figure}[t!]\centering
\includegraphics[width=0.5\textwidth]{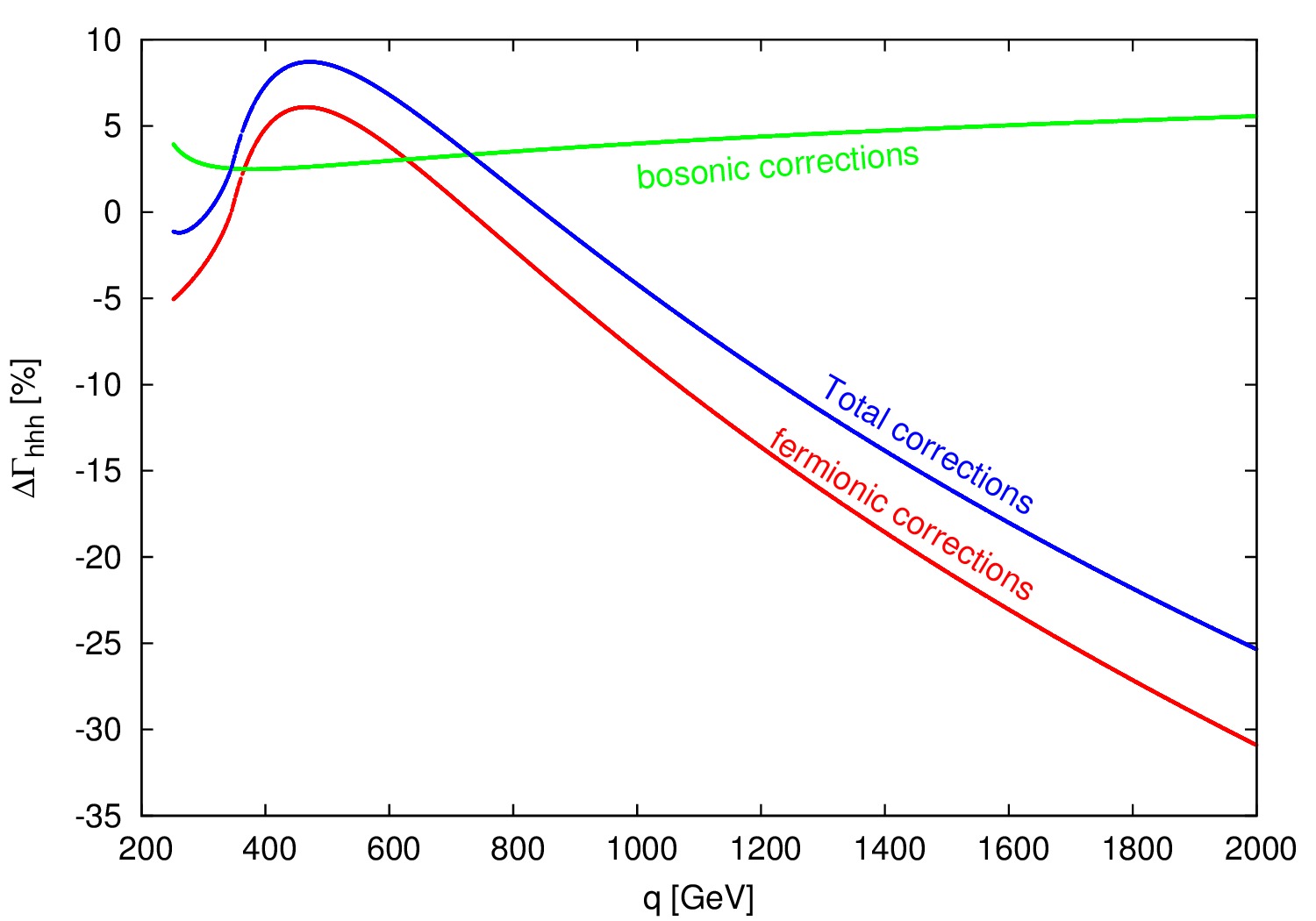}\includegraphics[width=0.5\textwidth]{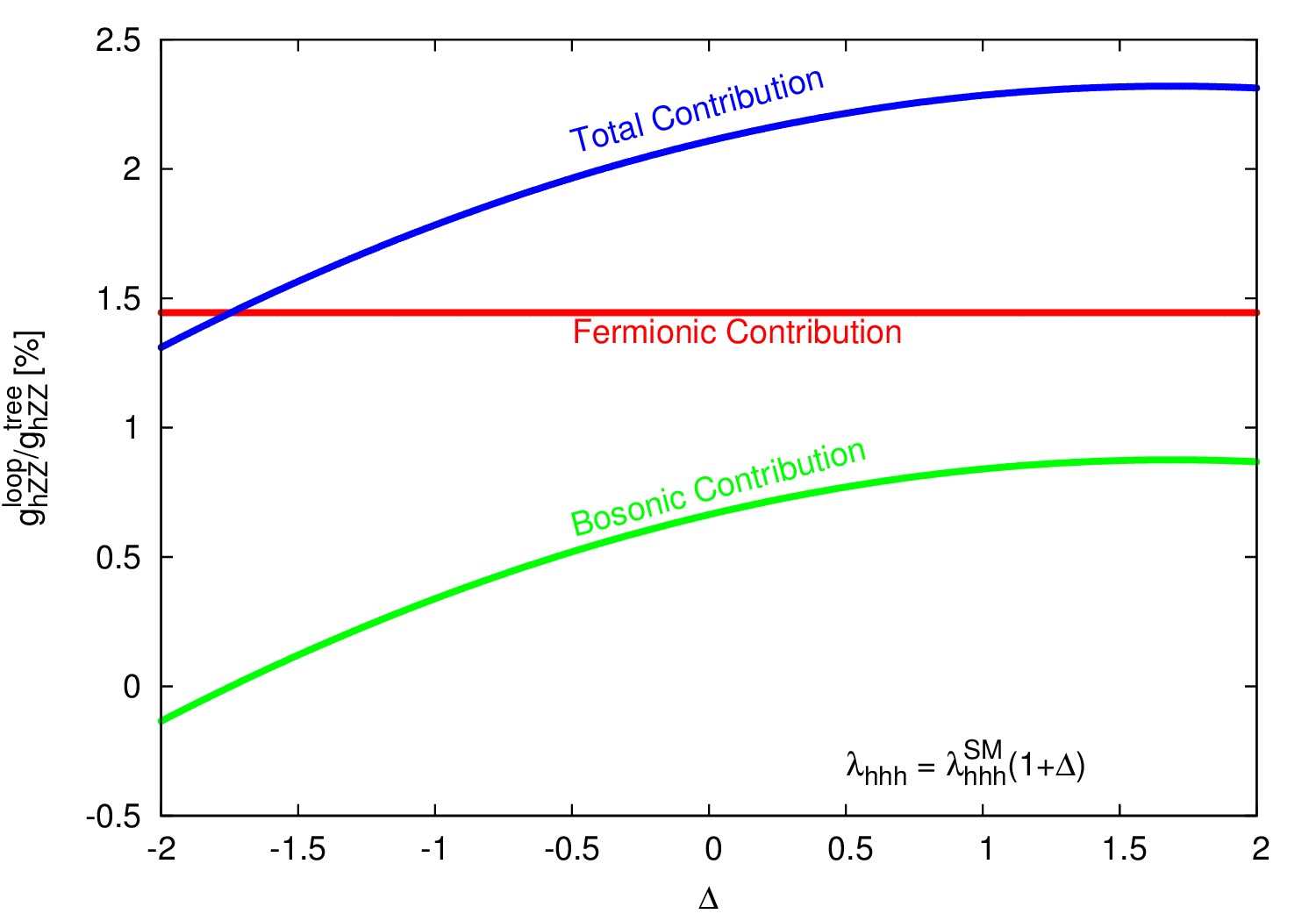}
  \caption{(Left) $\Delta\Gamma_{hhh}(SM)$ as a function of $h^*$ momentum $q$.
(Right) $\Delta\Gamma_{hZZ}(SM)$ as a function of $\Delta$ which is the size
  of deviation from SM triple coupling 
$\lambda_{hhh}=\lambda_{hhh}^{SM}(1+\Delta)$.
It is shown: the fermionic contribution, the bosonic one as well as 
the total contribution.}
\label{hhh-SM}
\end{figure}

In the SM, the dominant contribution to $\Delta\Gamma_{hhh}(SM)$
 comes from top quark loops
\cite{kanemura,penaranda}. We have computed the top contribution and shown
that it is in perfect agreement with Ref.~\cite{kanemura}. We have 
also isolated and evaluated the other SM contributions without fermions. 
 It turn out that this 
bosonic contribution is of the order of 5\% for large $q$. \\
In Fig.~(\ref{hhh-SM})(left), it is illustrated that the top contribution
 to $\Delta\Gamma_{hhh}(SM)$ is negative before the opening of 
$h^*\to t\bar{t}$ threshold and also for $q\geq 700$ GeV. 
It is clear that for large $q$, $\Delta\Gamma_{hhh}(SM)$ is 
dominated by top-quark contribution.\\
In Fig.~(\ref{hhh-SM})(right), we show the radiative corrections to $hZZ$ in
the SM. We present separately the fermionic corrections 
which are dominated by the top contributions and the bosonic contributions.
The total corrections to $hZZ$ is of the order of 2\%.
In this plot, we also shift the triple Higgs SM coupling $\lambda_{hhh}^{SM}$ by 
$\lambda_{hhh}^{SM} (1+\Delta)$, where $\Delta$ represent any deviation from
 SM coupling. As one can see from the green line, the sensitivity to $\Delta$
 is rather mild. Due to custodial symmetry, it is expected that $hWW$ coupling
 will enjoy similar effect as $hZZ$ and that is why we illustrate only the
 case of $hZZ$.
\subsection{IHDM case}
Here, we will show our numerical analysis for the triple coupling of the Higgs
in the IHDM taking into account: unitarity, perturbativity, false 
vacuum as well as vacuum stability constraints described above. 
We take the mass of the SM Higgs $m_h=125$ GeV and  the masses of the 
inert particles to be degenerate, i.e : $m_{H^\pm} = m_{H^0} = m_{A^0} =
m_{\Phi}$. For the other parameters, we perform the following scan:
\begin{eqnarray}
 100 \textrm{ GeV} \leq m_{\Phi} \leq 500 \textrm{ GeV} \nonumber \\
 -25 \times 10^5 \textrm{ (GeV)}^2 \leq \mu_2^2 \leq 9 \times 10^5 
\textrm{ (GeV)}^2 \nonumber \\
  0 <  \lambda_2 \leq \frac{4 \pi}{3}
 \label{para.range}
\end{eqnarray}

\begin{figure}[t!]\centering
\includegraphics[width=0.5\textwidth]{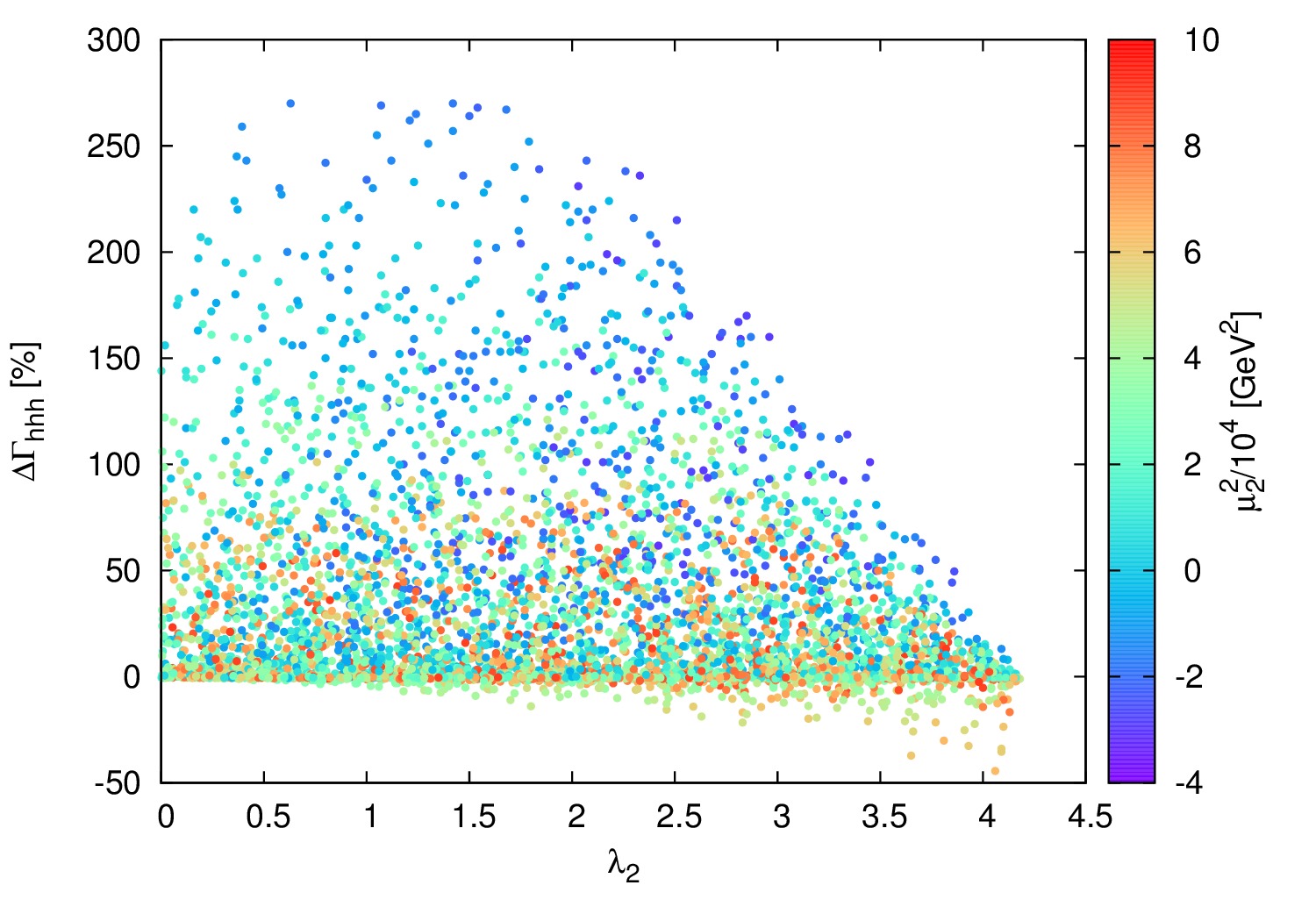}\includegraphics[width=0.5\textwidth]{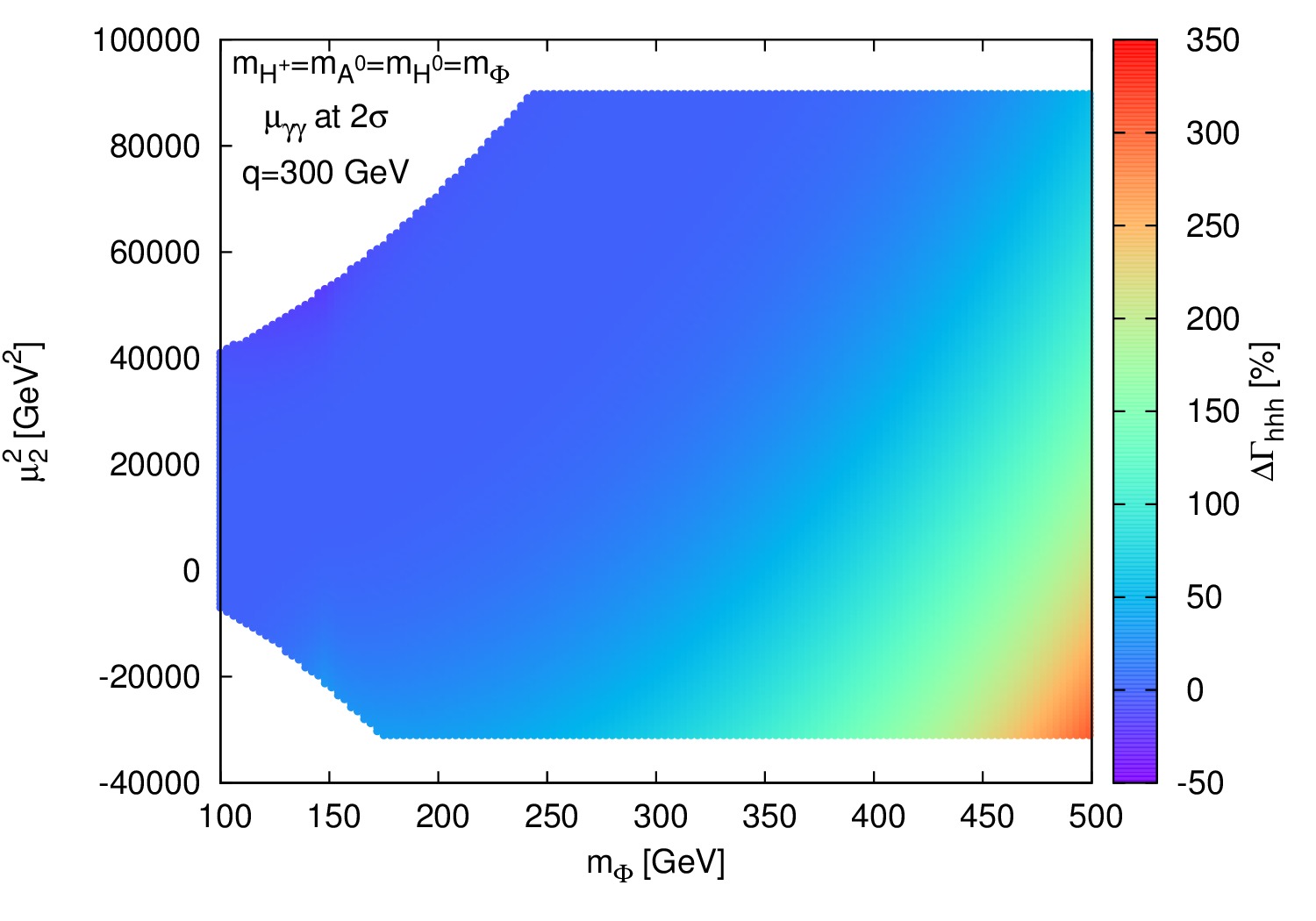}
\caption{Left: $\Delta\Gamma_{hhh}(IHDM)$ as a function of $\lambda_2$.
Right: Scatter plot for $\Delta \Gamma_{hhh}$ in the plan $(m_{\Phi}, 
\mu_2^2)$ for $q=300$ GeV, Left column show the size of the corrections.
$m_\Phi$ and  $\mu_2^2$ are scanned as in eq.~(\ref{para.range}).
 }
\label{hhh-IHD1}
\end{figure}
We plot in Fig.~(\ref{hhh-IHD1})(left) the relative corrections to the 
triple coupling  as a function of $\lambda_2$. The theoretical constraints put a limit on 
$\lambda_2$ parameter which is $\lambda_2 \leq \frac{4 \pi}{3}$. 
One can see from Fig.~\ref{hhh-IHD1} that the corrections are maximized for 
$\lambda_2 \leq 2$ and  decrease for $\lambda_2 > 2$. In our following
analysis, We will take $\lambda_2 = 2$ in order to maximize the 
effect from $\lambda_2$.\\
In Fig.~(\ref{hhh-IHD1})(right), we plot the relative corrections to the triple
coupling $hhh$  in the plane ($m_\Phi$, $\mu_2^2$) for a fixed 
$q=300$ GeV and $\lambda_2=2$. One can see that the corrections 
are very important in a large part of the
parameter space with an enhancement up to $280 \%$ for large 
values of $m_{\Phi}$ and negative $\mu_2^2$. 
Furthermore, these
corrections are increasing, for a fixed value of $m_\Phi$, 
while $\mu_2^2$ is decreasing. The maximum of the corrections is
reached for $\mu_2^2 \approx -30000 \textrm{ (GeV)}^2$. \\
Moderate or very small corrections which can be in 
the range $[-50,50] \%$  are also possible for large area of   
parameter space with  low $m_\Phi \leq 300 \textrm{ GeV}$ and any positive 
$\mu_2^2$. It is also important to note that LHC constraint from diphoton
 at the 2$\sigma$ level exclude light charged Higgs 
$100< m_{H^\pm} <175$ GeV and negative $\mu_2^2$: left-down corner of the
 scatter plot.
\begin{figure}[t!]\centering
\includegraphics[width=0.5\textwidth]{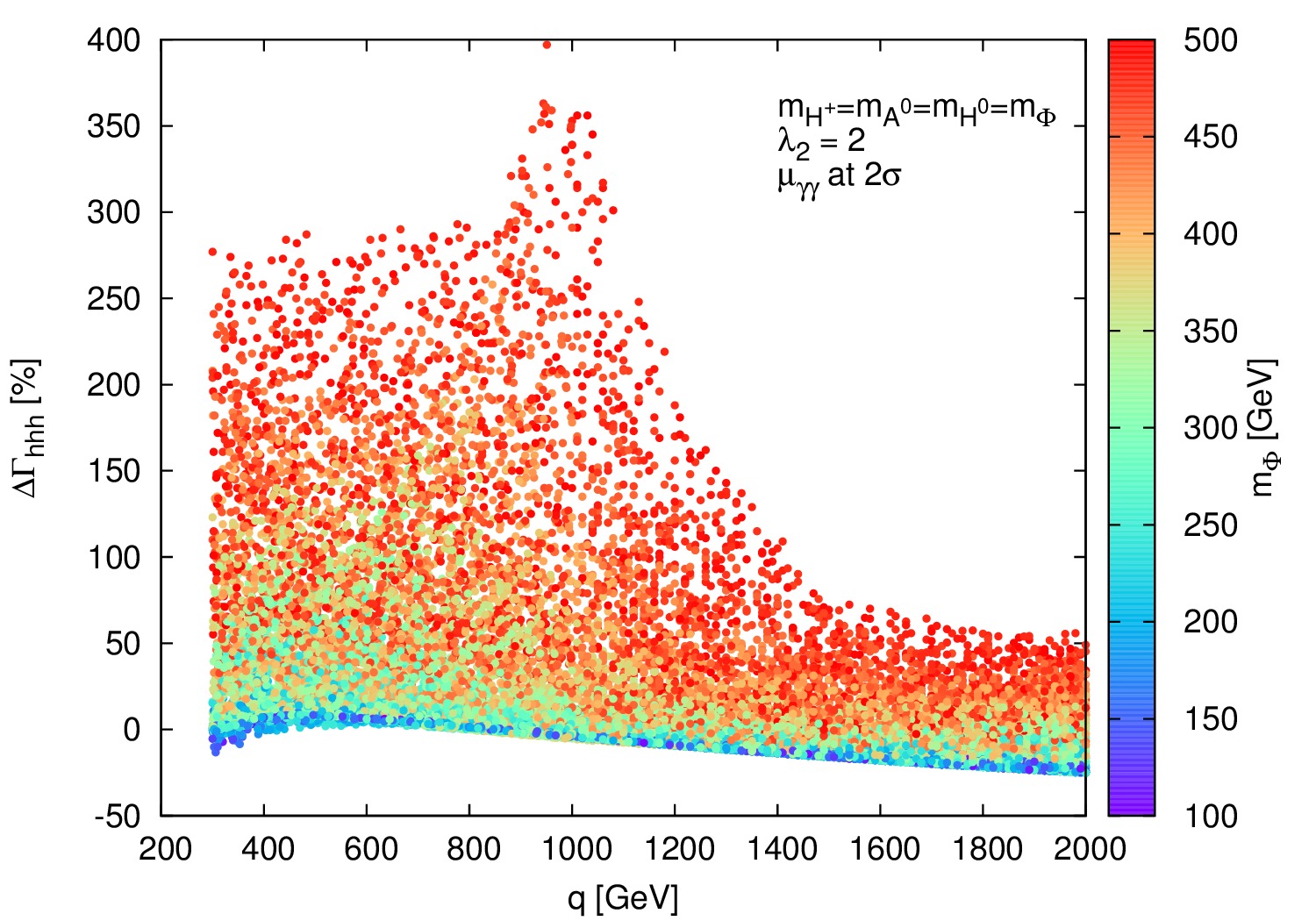}\includegraphics[width=0.5\textwidth]{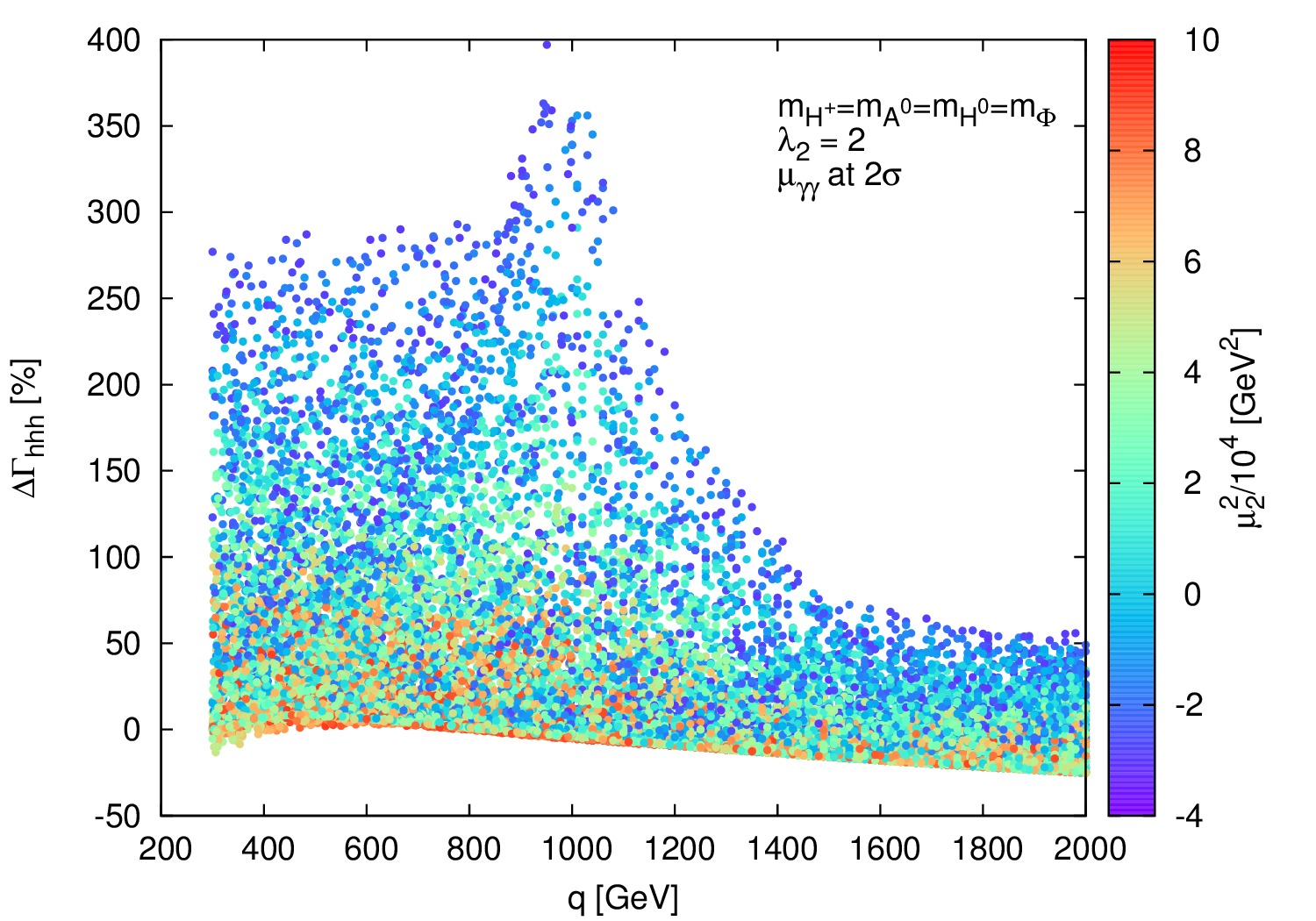}
\caption{$\Delta\Gamma_{hhh}(IHDM)$ as a function of $q$ and where $m_\Phi$ and
  $\mu_2^2$ are scanned as in eq.~(\ref{para.range}). The left column shows
  the values of $m_\Phi$ (left panel) and $\mu_2^2$ (right panel)}
\label{hhh-IHD2}
\end{figure}
In Fig.~\ref{hhh-IHD2} we show the relative corrections
$\Delta\Gamma_{hhh}(IHDM)$ as a function of the momentum of the 
off-shell decaying Higgs $q$ and for  $m_\Phi$ and
  $\mu_2^2$ as shown in eq.~(\ref{para.range}). It is clear that for 
low $100< m_{\Phi}<200 $ GeV, the corrections are small like in the 
SM case except at the threshold regions where we have 
$\Delta\Gamma_{hhh} \sim -40 \%$ due to the opening of $h^*\to \Phi\Phi$. 
For large $m_{\Phi}$ these corrections could be extremely large
 exceeding 100\% in large area of parameter space. The corrections are
 amplified by the opening of the threshold channel $h^* \to \Phi \Phi$.
This is visible on the left panel of Fig.~\ref{hhh-IHD2} where we can see a 
kink for $q=1000$ GeV which correspond to threshold effect 
$h^* \to \Phi \Phi$ with $m_{\Phi}\approx 500$ GeV. As it is shown, 
negative values for $\mu_2^2$ gives large corrections to the triple Higgs 
coupling. This is because in our assumption of taking 
 degenerate Higges $m_{H^0}=m_{A^0}=m_{H^\pm}=m_{\Phi}$
 one can show that $\lambda_4=\lambda_5=0$, $\lambda_3=\frac{2}{v^2}
 (m_{H^\pm}^2-\mu_2^2)$ and therefore the triple coupling are given by
\begin{eqnarray}
hHH&=& \lambda_Lv=\frac{2}{v}
 (m_{\Phi}^2-\mu_2^2)\nonumber\\
hAA&=& \lambda_A v=\frac{2}{v}
 (m_{\Phi}^2-\mu_2^2)\nonumber\\
hH^+H^-&=& \lambda_3\,v=\frac{2}{v}
(m_{\Phi}^2-\mu_2^2)
\end{eqnarray}
It is clear that those couplings gets  stronger for negative $\mu_2^2$.
\begin{figure}[t!]\centering
\includegraphics[width=0.5\textwidth]{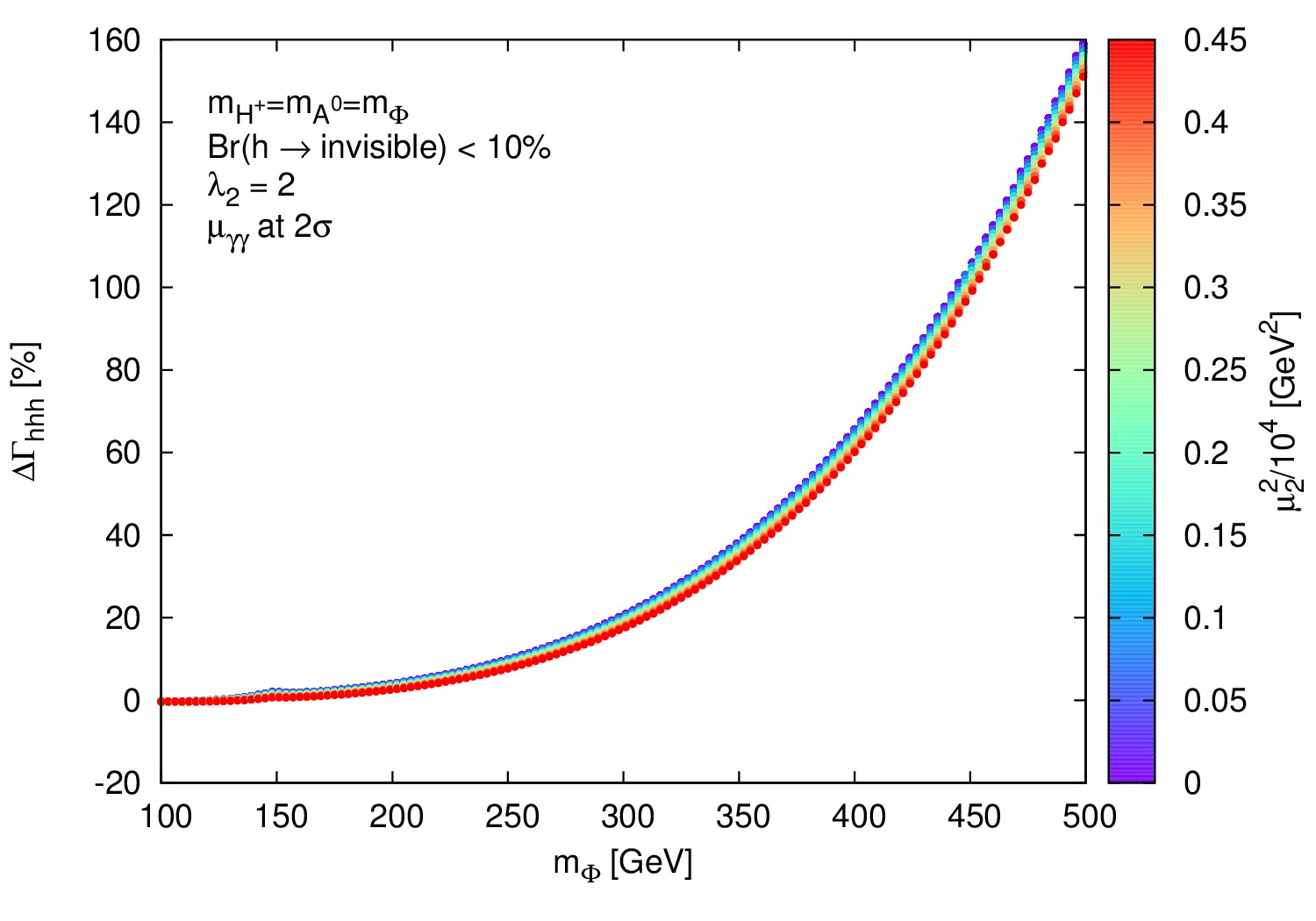}\includegraphics[width=0.5\textwidth]{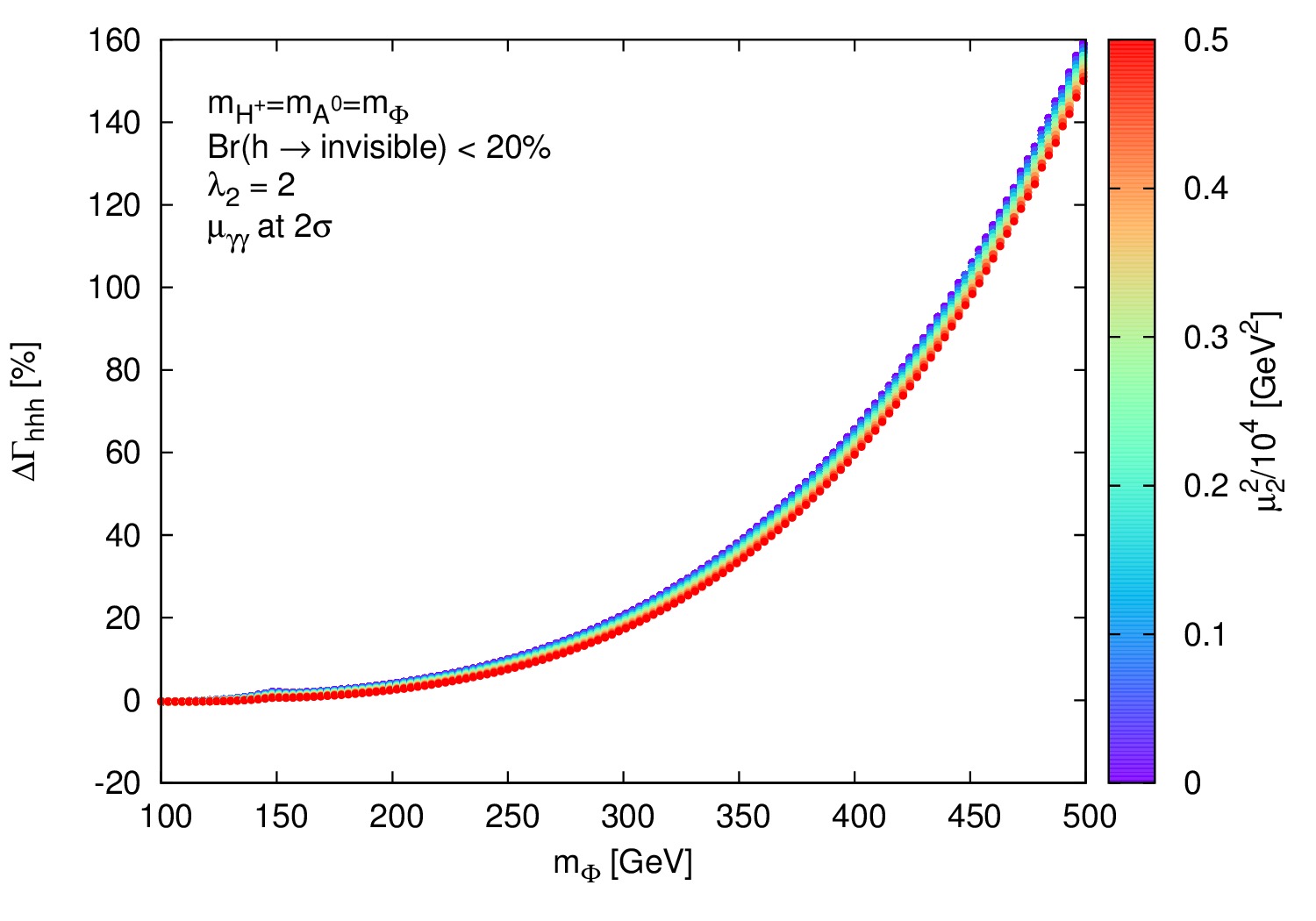}
 \caption{$\Delta \Gamma_{hhh}$ as a function of $m_{\Phi}$ for 
$q=300 \textrm{ GeV}$ with $\mu_2^2$ in the range given by 
the Eq. (\ref{para.range}) and invisible decay of the Higgs ($h\to HH$) 
is open. Left panel with dark matter constraint 
which restrict $|\lambda_L|$ to be in the range $|\lambda_L|<0.02$. 
Right panel without dark matter constraint.}   
\label{hhh-IHD3}
\end{figure}
We now examine the effect of the radiative corrections on the triple coupling 
in the case where the invisible decay $h\to HH$ is open.
This is illustrated in Fig.~\ref{hhh-IHD3}(left) and right. 
In Fig.~\ref{hhh-IHD3}(left) we impose
both $|\lambda_L|<0.02$ required by dark matter constraints 
 as well as  best fit limit on the invisible branching ratio 
$Br(h\to invisible)\leq 10\%$(left) and $Br(h\to invisible)\leq 20\%$(right)  
 \cite{invisible}.
It is clear from left panel that with dark matter constraint, the size of the
corrections and the range of $\mu_2^2$  are smaller than in the previous case 
where the invisible decay was closed. The large corrections observed 
for high $m_{\Phi}$ are mainly due to the charged Higgs loops. 

%
\begin{figure}[t!]\centering
\includegraphics[width=0.5\textwidth]{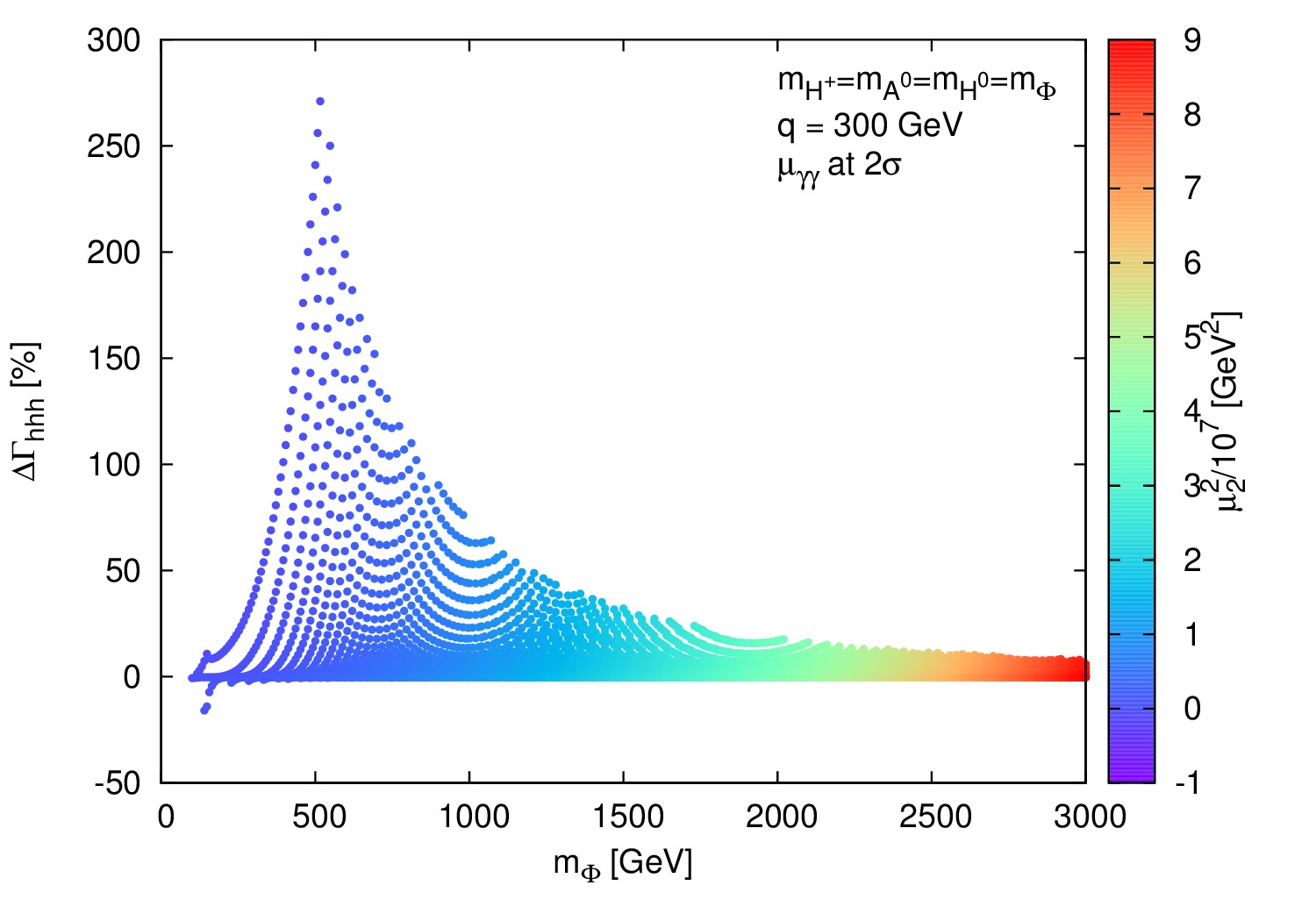}\includegraphics[width=0.5\textwidth]{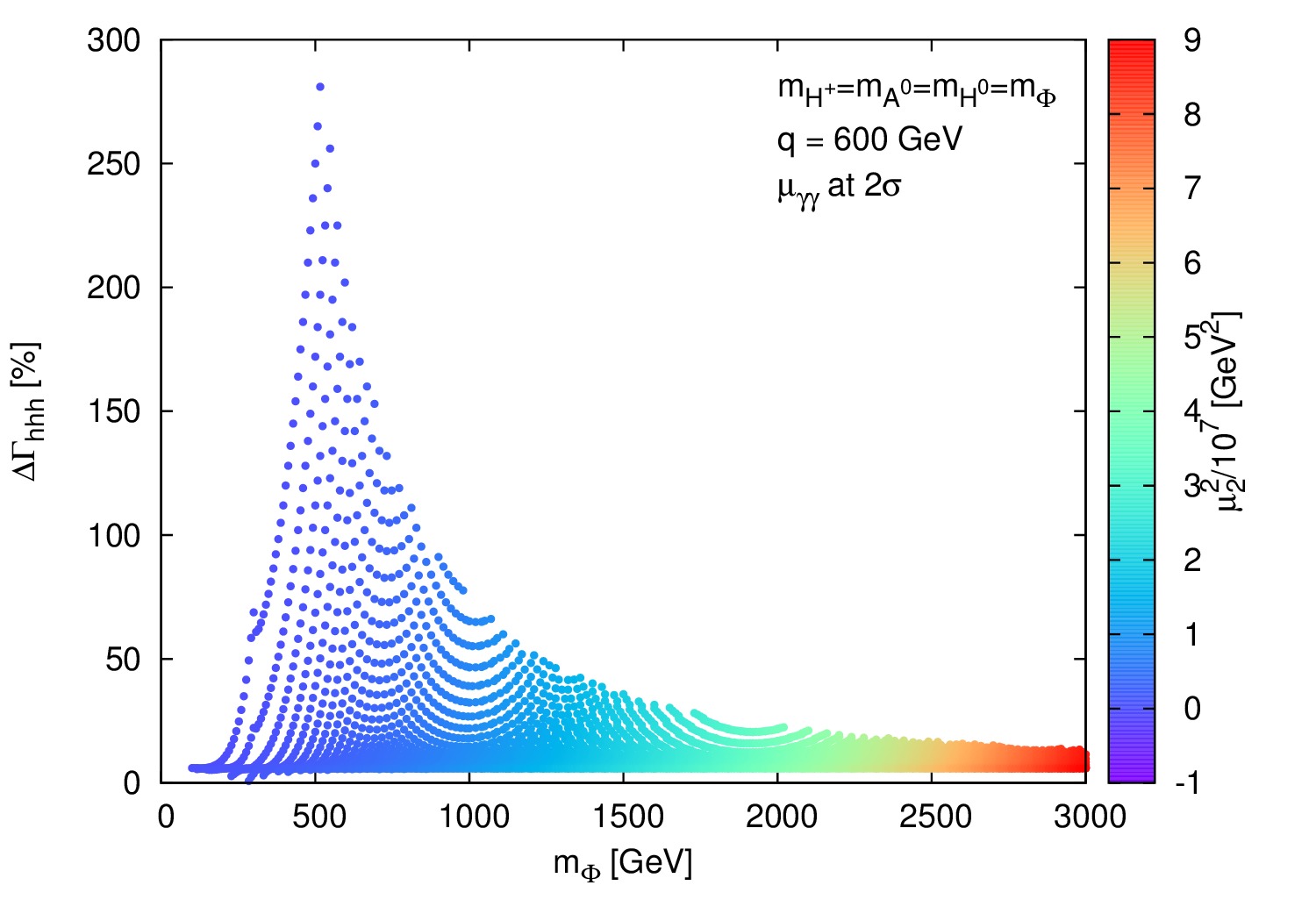}
\caption{Left: $\Delta\Gamma_{hhh}(IHDM)$ as a function of $m_{\Phi}$ for
  fixed $\lambda_2 = 2$ and $q=300$ GeV (left), while $q=600$ GeV in the right.
 Left column show the range of  $\mu_2^2$.}
\label{hhh-decoupling}
\end{figure}
In order to exibit the decoupling behavior on the triple Higgs coupling, 
We increase both the range of $\mu^2_{2}$ to be  $[-10^6,10^7]$ 
$GeV^2$ as well as the range of $m_{\Phi}\in [0.1,3]$ TeV.
We see that the decoupling effect occurs when appropriate
combinations of the involved parameters are taken large compared to the
electroweak scale. We find that $\Delta \Gamma_{hhh}$ reaches its maximum for
$m_\Phi\approx$ 500 GeV and decrease to SM value for large $m_\Phi$.

\begin{figure}[t!]\centering
\includegraphics[width=0.5\textwidth]{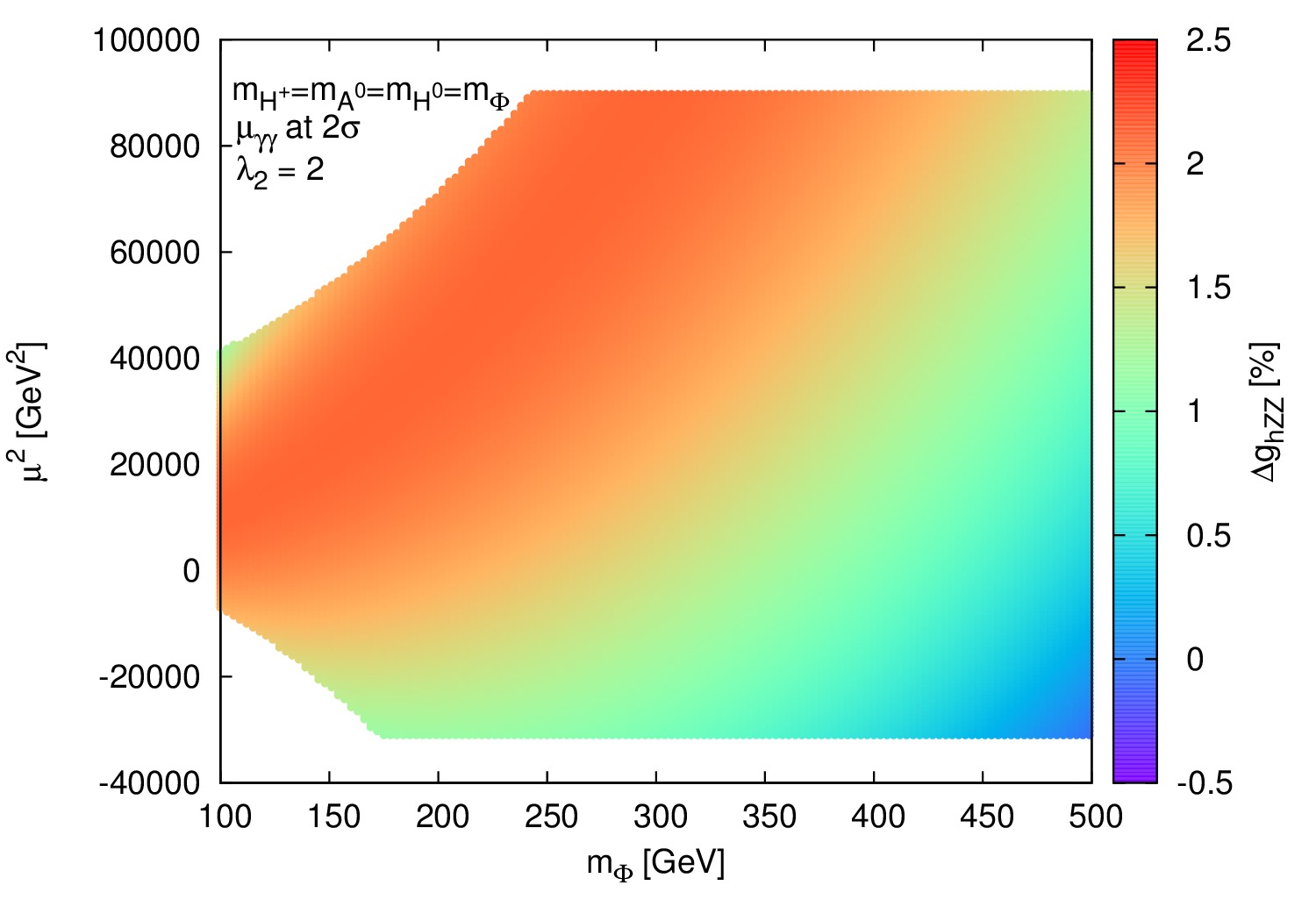}\includegraphics[width=0.5\textwidth]{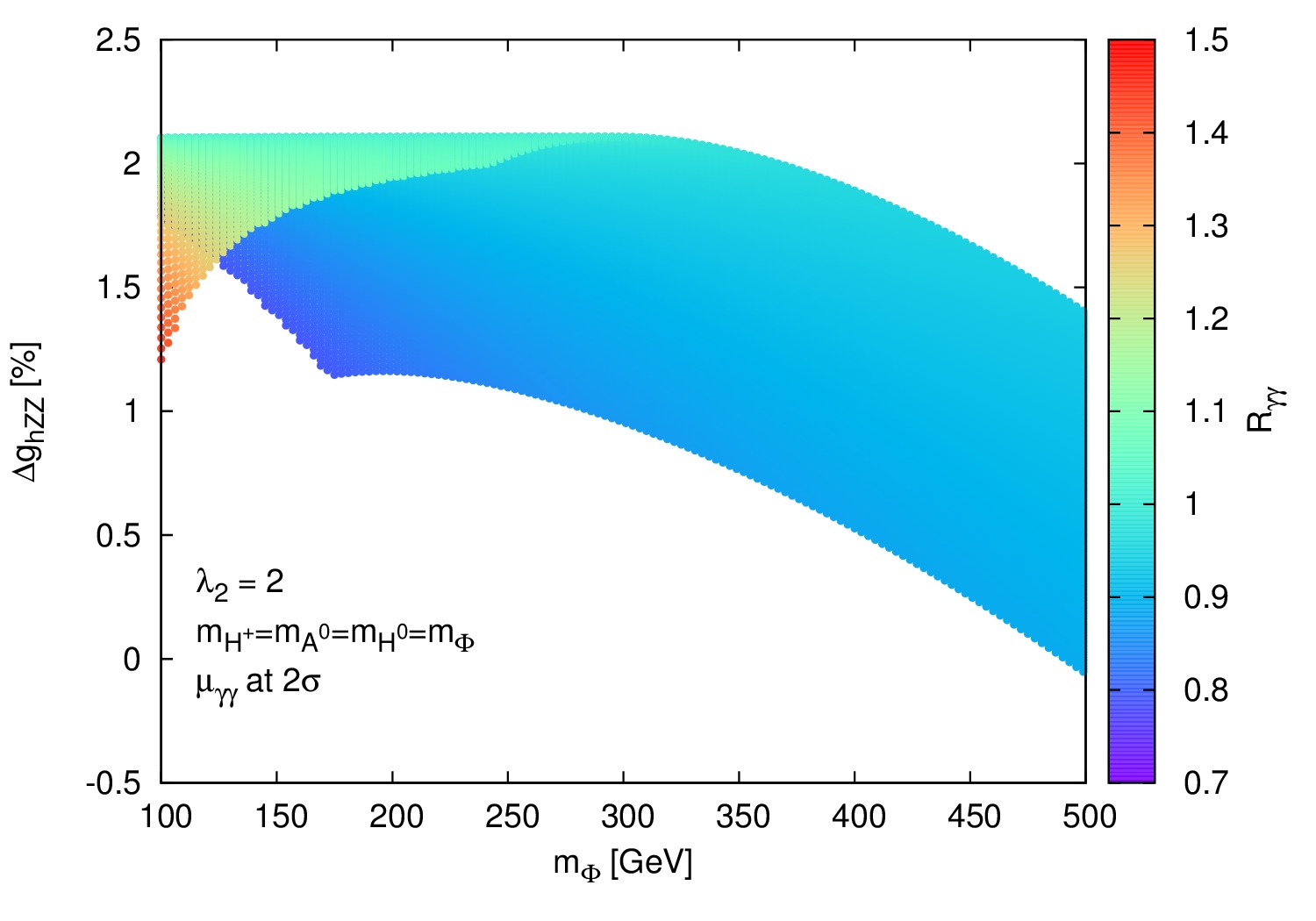}
 \caption{Left: Scatter plot for $\Delta \Gamma_{hZZ}$ in the plan 
$(m_{\Phi},\mu_2^2)$ with $\lambda_2=2$, left column represent the size of the
   corrections. Right:  $\Delta \Gamma_{hZZ}$ as function of $m_{\Phi}$ with
   $\mu_2^2$ in the same range as in the left panel. Left column show the size
 of $R_{\gamma\gamma}$.}   
\label{hzz-IHDM}
\end{figure}

In Fig.~(\ref{hzz-IHDM}) we illustrate the IHDM effect on $hZZ$ coupling.
Similar to the triple Higgs coupling, we fix $\lambda_2=2$ and scan over
$\mu_2^2$ and $m_{\Phi}=m_H=m_A=m_{H\pm}$ as in eq.~(\ref{para.range}).
In Fig.~(\ref{hzz-IHDM})(left) we show scatter plot for $\Delta \Gamma_{hZZ}$ 
 in the plan  $(\mu_2^2,m_{\Phi})$, contrarily to the triple coupling $hhh$
 the effect are rather small of the same size as in the SM case. In 
Fig.~(\ref{hzz-IHDM})(right) we show the corrections to $hZZ$ coupling 
and also the value of the corresponding $R_{\gamma\gamma}$ within 2$\sigma$
range.
 
\section{Radiative corrections to $e^+ e^- \to hhZ$ in the IHDM}
\subsection{$e^+ e^- \to hhZ$ in SM}
\begin{figure}[ptb]
\begin{center}
\input{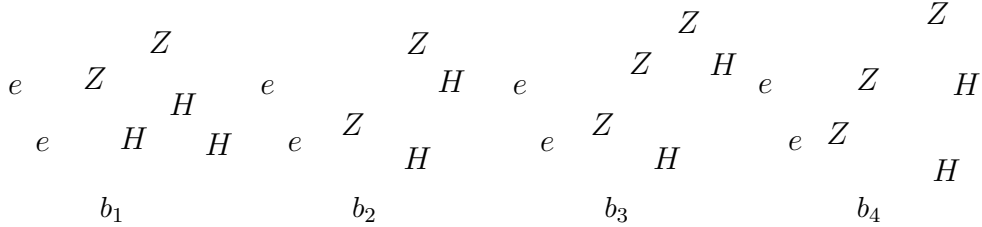}
\end{center}
\vspace{-13.8cm}
\caption{Feynman diagrams contributing to $e^+e^- \to Zhh$ at 
the tree level  in SM}
\label{fdhhh}
\end{figure}
At $e^+e^-$ LC, the triple Higgs coupling can be probed by double 
Higgsstrahlung process $e^+e^-\to hhZ$ depicted in 
 Fig.~(\ref{fdhhh}) and WW fusion process 
$e^{+}e^-\to WW^{*}\nu\bar{\nu}\to hh\nu\bar{\nu}$.
Note that $e^+e^-\to hhZ$ process 
arise in $s-$channel only and hence its cross section can be probed 
more efficiently at low
energies above the threshold (typically between $350$ GeV and $500$ GeV). 
While, at high energies, for $\sqrt{s} \gtrsim 700 \textrm{ GeV}$, 
the trilinear Higgs-coupling is better probed through the process  
$e^+ e^- \to \nu \bar{\nu} hh$ assuming the SM. 
In Fig.~(\ref{fdhhh}) we show the Feynman diagrams contributing to
 $e^+e^-\to hhZ$. When extracting the triple Higgs coupling $hhh$ from this
process only diagram Fig.~(\ref{fdhhh})($b_1$) is concerned, the other
diagrams $b_{2,3,4}$ are considered as a background. \\
We illustrate in Fig.~(\ref{fdhhh}) the tree level cross section for 
$e^+e^-\to hhZ$ as a function of $\Delta$ for center of mass energy 500 
GeV and 800 GeV, where $\Delta$ is the shift of the SM triple Higgs coupling 
$\lambda_{hhh}=\lambda_{hhh}^{SM} (1+\Delta)$. It is clear that for 
$\Delta >0$ the cross section is enhanced while for $\Delta <0$ the cross section is reduced with respect to SM value.
\begin{figure}[t!]\centering
\includegraphics[width=0.6\textwidth]{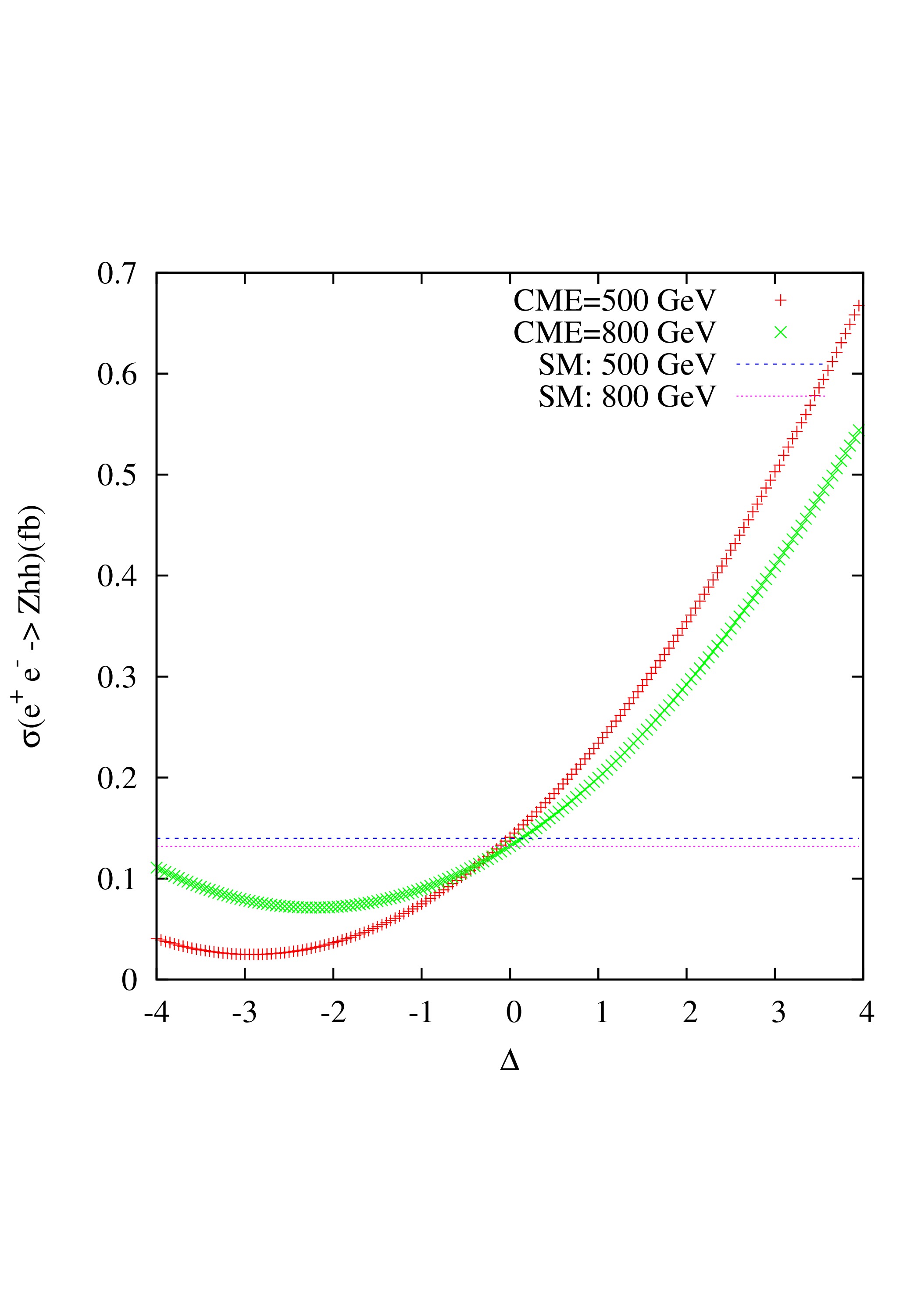}
\vspace{-2cm}
 \caption{Cross section of $e^+e^-\to Zhh$ as a function of $\Delta$ where 
$\lambda_{hhh}=\lambda_{hhh}^{SM} (1+\Delta)$ for $\sqrt{s}=500, 800$ GeV.  }
 \label{zhhtree}
\end{figure}
\subsection{One-loop Corrections to $e^+e^-\to Zhh$}
We study in this section the effects of the one-loop radiative 
corrections to the Higgs trilinear self-coupling calculated in the 
previous section on the double Higgs-Strahlung process via $Z$ boson exchange.
In the context of the SM, the
$\cal{O}(\alpha)$ electroweak corrections have been studied in 
 \cite{F.Boudjema} and it was found that these corrections are of 
the order $10 \%$. 
However, these loop effects can be very large in beyond SM enhancing the total 
cross section by about $2$ orders of magnitude in particular in models with 
extended Higgs sector. As outlined above,
at the tree level, $e^+e^-\to hhZ$ have four diagrams as depicted in
Fig.~(\ref{fdhhh}). Only the first diagram contribute to the signal while the 
others are considered as a background.
In our analysis, we include one-loop correction only to the
 triple Higgs coupling $hhh$ which is expected to give sizeable contribution.
Only this correction contribute to the signal. 
Therefore, we did not include corrections to the initial state vertex
$e^+e^-Z$, to the self energies of Z-Z and $\gamma-Z$ mixing, to the 
 $hZZ$ coupling and also we did not include the initial state radiation. 
In fact these corrections are well known in the SM and 
are not expected to deviate that much in the IHDM as we already show in the
previous section for $hZZ$ coupling. Moreover, we will not include corrections
to $e^+e^-\to hhZ$ coming from boxes and pentagon diagrams.
The one-loop amplitude can be written as follow:
\begin{eqnarray}
 {\mathcal{M}} = {\mathcal{M}}_{\textrm{tree}} + \mathcal{M}_{\textrm{loop}}
\end{eqnarray}
The squared amplitude at the one-loop level is then :
\begin{eqnarray}
 |{\mathcal{M}}|^2 = |{\mathcal{M}}_{\textrm{tree}}|^2 + 
2 \textrm{Re}\{{\mathcal{M}}_{\textrm{tree}}^*\mathcal{M}_{\textrm{loop}}\} 
+ {\cal{O}}(\alpha^2)
\label{squar.amp}
 \end{eqnarray}
Thus, the cross section is written as:
\begin{eqnarray}
 \sigma = \frac{1}{(2 \pi)^2} \int \prod_{k=1}^{3} 
\frac{d^3 \textbf{p}_k}{2 E_k} \delta^{(4)} (q_1 + q_2 - p_1 - p_2 - p_3) 
 \sum_{\textrm{spin, pola.}} |{\mathcal{M}}|^2
\end{eqnarray}
where $q_1$ and $q_2$ are the four momentum of the incoming 
electron and positron, $p_1, p_2 \textrm{ and } p_3$ are the 
four momentum of the outgoing particles,
and the factor $\frac{1}{(2 \pi)^2}$ arises from the flux of the 
initial particles.

For our studies, we define the ratio $\Delta \sigma$ by :
\begin{eqnarray}
 \Delta \sigma = \frac{\sigma_{\textrm{total}} - 
\sigma_{\textrm{tree}}}{\sigma_{\textrm{tree}}} = 
\frac{\sigma_{\textrm{loop}}}{\sigma_{\textrm{tree}}}
 \label{sigma.ratio}
\end{eqnarray}
Where $\sigma_{\textrm{total}} = \sigma_{\textrm{tree}} + \sigma_{\textrm{loop}}$.
This ratio measure the relative correction of the IHDM to the cross section
 with respect to the tree level result. 

\begin{figure}[t!]\centering
\includegraphics[width=0.5\textwidth]{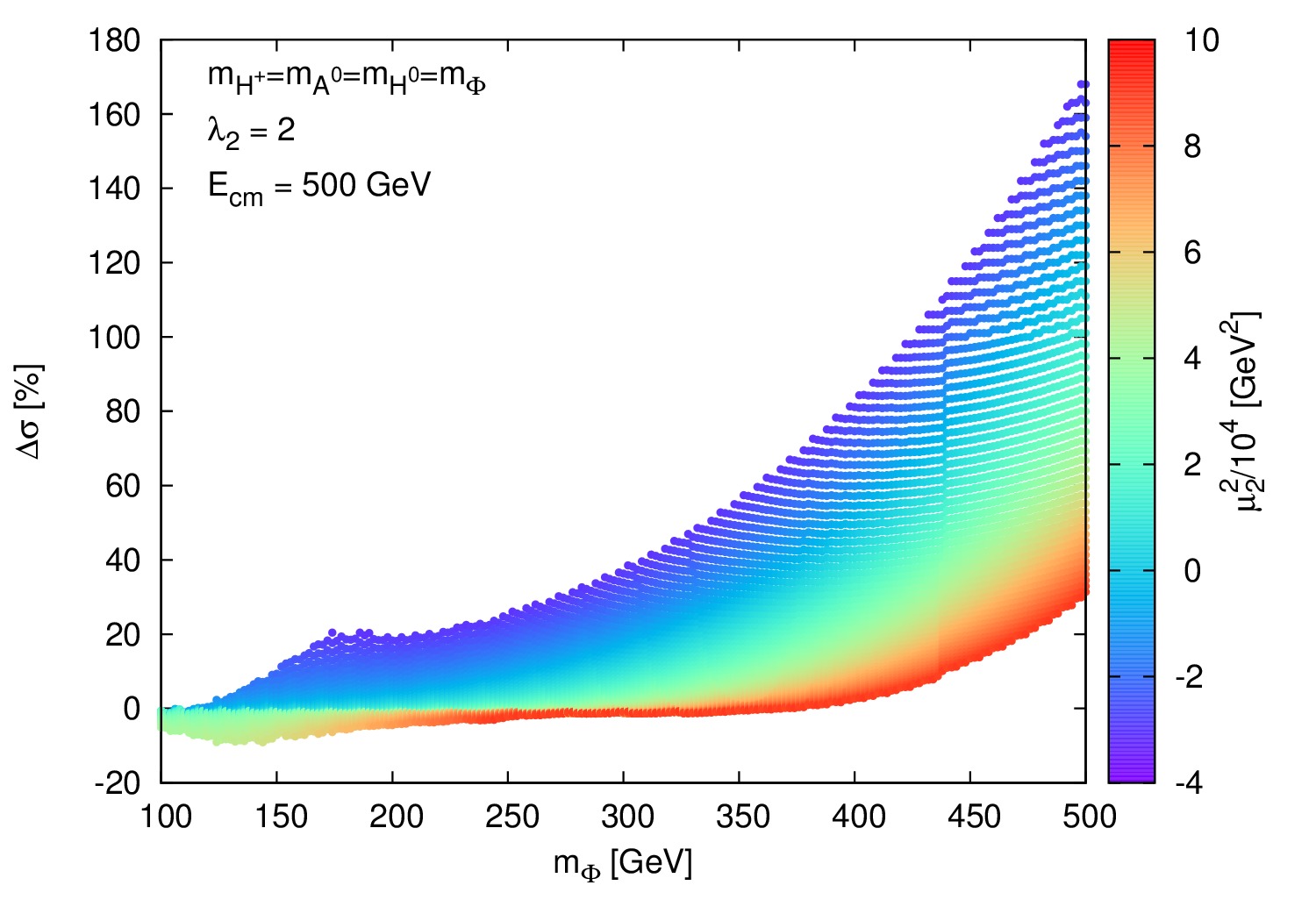}\includegraphics[width=0.5\textwidth]{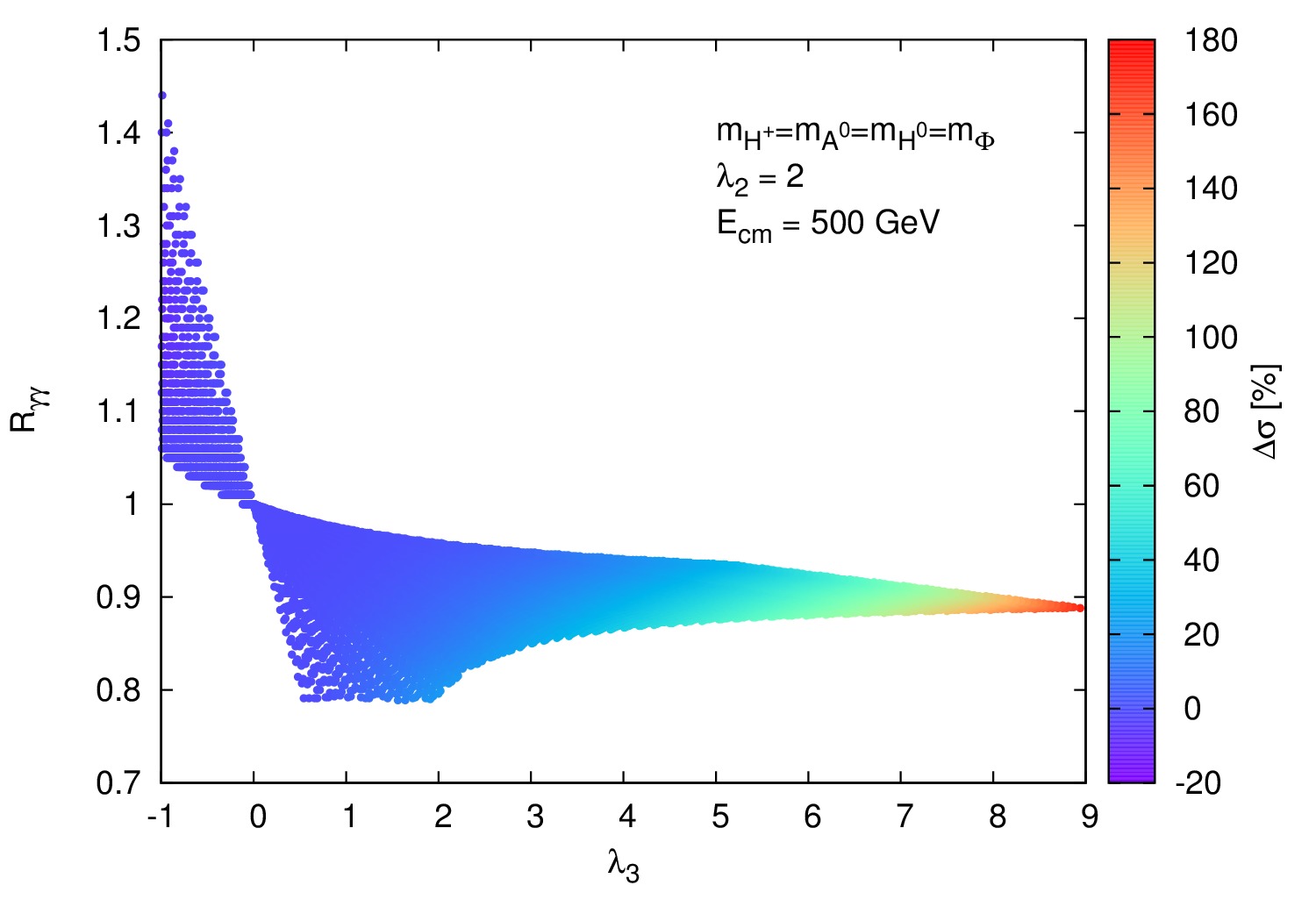}
 \caption{(left): $\Delta \sigma$ as a function of  
$m_\Phi$ for $\sqrt{s} = 500 \textrm{ GeV}$ $\mu_2^2$ values are shown in the
   right column, 
(right) 
$R_{\gamma\gamma}$ as a function of $\lambda_3$ and the right column shows 
$\Delta\sigma$ in \% .}
 \label{Sec-IDH1}
\end{figure}

As stated before, in our analysis we will take into account the theoretical
and experimental constraints discussed in the second section assuming 
the parameters to rely in the range given by eq.~(\ref{para.range}). 
The phase space and evaluation of the one-loop squared amplitude 
has been performed with  FormCalc \cite{FA2} with the help of LoopTools 
to evaluate numerically the one-loop scalar integrals. 
We have used the same on-shell renormalization scheme explained in the 
previous section. \\
In Fig.~\ref{Sec-IDH1}, we have plotted the ratio $\Delta \sigma$ 
versus $m_\Phi$ for center of mass energy $\sqrt{s} = 500 \textrm{ GeV}$. 
We assume again that: $\lambda_2=2$, the dark Higges to be degenerate 
$m_H=m_A=m_{H\pm}=m_{\Phi}$ and perform a scan over $\mu_2^2$ and $m_{\Phi}$.
From this plot one can see that the corrections can reach 
$160 \%$ for high dark Higgs masses $m_\Phi \approx 500 \textrm{ GeV}$ 
and are negative for low dark Higgs masses 
$100 \textrm{ GeV} \leq m_\Phi \leq 270 \textrm{ GeV}$ 
depending on the value of $\mu_2^2$. 
One can see that the suppression of the total 
cross section can reach $-15 \%$ for $m_\Phi = 140 \textrm{ GeV}$ while the 
enhancement is predominant in most part of the parameter space. \\
To understand this, let us remind first 
that in large area of parameter space the correction to the 
triple Higgs coupling $\Delta\Gamma_{hhh}$  
is positive (see section IV Fig.~(\ref{hhh-IHD1})). 
Moreover, according to the plot Fig.~(\ref{zhhtree}),
if this correction is positive this lead to an enhancement 
of the total cross section and vice-versa.\\
This explain that in most of the case, the corrections to the cross section
are  positive and confirm that the behavior of the
$\Delta \sigma$ is consistent with our analysis
concerning the trilinear Higgs self coupling in the IHDM. We stress that 
the enhancement of $\Delta \sigma$ is 
observed in a large part of the parameter space and 
can exceed $100 \%$ only in the high mass region $m_\Phi 
\geq 400 \textrm{ GeV}$ for $\mu_2^2 < 0$.

We plot in Fig.~\ref{Sec-IDH1}(right) the ratio $R_{\gamma\gamma}$ as a 
function of $\lambda_3$ and showing the relative corrections in the left column.
For small values of $-1<\lambda_3<2$ (low $m_{\Phi}$) 
the corrections are quite small. For $\lambda_3 > 4$ (high $m_{\Phi}$),
 $\Delta\sigma$ becomes more important and exceed 100\%, 
this region corresponds to 
$R_{\gamma\gamma} \approx 0.9 \pm 0.02$. Note that our results concerning
 $R_{\gamma\gamma}$ are in agreement with the results of \cite{htogaga}.

\section{Conclusion}
We have computed the radiative corrections to triple Higgs coupling $hhh$, 
$hZZ$ coupling as well as $e^+e^- \to Zhh$ in the framework of inert Higgs
doublet model taking into account theoretical and experimental constraint 
on the parameter space of the model. The calculation was done in the Feynman
gauge using dimensional regularization in the on-shell scheme. In the SM 
it is known that the top contribution to $hhh$ coupling is of the order 10\%,
 we found that the bosonic contribution is somehow significant and goes up to 
5\%. In the IHDM, we found that
the total radiative corrections to the triple Higgs coupling could be
substantial exceeding 100\% for heavy dark Higgs masses $m_H, m_A$ and
$m_{H\pm}$. We also show that the corrections to $hhh$ are decoupling 
for large $m_{\Phi}$ and large $\mu_2^2$.
In the case of $hZZ$ coupling the effect is rather mild and do not
exceed 2.5\%.  We also evaluate radiative corrections to the double Higgs 
strahlung process $e^+e^-\to Zhh$ by looking only to the correction to the
diagram that contribute to the signal i.e the triple coupling $hhh$. 
We have shown that the correction are also very important. 
In general, the size of the loop effects,  typically large, 
makes their proper inclusion in phenomenological analyses for future
$e^+e^-$ LC indispensable.

\section*{Acknowledgments}

A.A would like to thank NCTS for financial support where part of this work has
been done. This work was supported by the Moroccan Ministry of Higher
Education and Scientific Research MESRSFC and  CNRST: "Projet dans les domaines prioritaires
de la recherche scientifique et du d\'eveloppement technologique": PPR/2015/6.
A.J would like to thank ICTP-Trieste for financial support 
during his stay where part of this work has been done.

\section*{Appendix}
\subsection*{The Triple Higgs Coupling in the IHDM : Analytical Results}
In this appendix, we present the analytical expression of the Higgs 
triple coupling at the one-loop 
order with the contribution of the inert scalars only. We use the 
Feynman diagrammatic method. 
The Feynman diagrams contributing to this process are shown in 
Fig. ~\ref{FD}. 
By using the dimensional regularization, the amplitude is given by:
\begin{eqnarray}
\Gamma_{hhh}^{loop} (q^2,m_\Phi^2)&&=\frac{\lambda_3^{2}m_{W}s_{W}}{8e\pi^2}
\biggl( B_{0}(q^2,m_{H\pm}^{2},m_{H\pm}^{2})+2B_{0}(m_{h}^{2},m_{H\pm}^{2},m_{H\pm}) \nonumber \\
&&+\frac{2 \lambda_3 m_{W}^2s_{W}^2}{ \pi \alpha } C_{0}(q^2,m_{h}^{2},m_{h}^{2},m_{H\pm}^{2},m_{H\pm}^{2},m_{H\pm}^{2})\biggr) \nonumber \\
&&+\frac{(\lambda_3+\lambda_4+\lambda_5)^{2} m_{W}s_{W}}{16 e \pi^2}\biggl( B_{0}(q^2,m_{H^0}^{2},m_{H^0}^{2})+2B_{0}(m_{h}^{2},m_{H^0}^{2},m_{H^0}^{2}) \nonumber\\
&& \times\frac{2(\lambda_3+\lambda_4+\lambda_5) m_{W}^2s_{W}^2}{\pi \alpha}C_{0}(q^2,m_{h}^{2},m_{h}^{2},m_{H^0}^{2},m_{H^0}^{2},m_{H^0}^{2})\biggr) \nonumber \\
&&+\frac{(\lambda_3+\lambda_4-\lambda_5)^{2} m_{W}s_{W}}{16e\pi^2}\biggl(B_{0}(q^2,m_{A^0}^{2},m_{A^0}^{2})+2B_{0}(m_{h}^{2},m_{A^0}^{2},m_{A^0}^{2}) \nonumber \\
&&\times\frac{2(\lambda_3+\lambda_4-\lambda_5) m_{W}^2s_{W}^2}{\pi \alpha}C_{0}(q^2,m_{h}^{2},m_{h}^{2},m_{A^0}^{2},m_{A^0}^{2},m_{A^0}^{2})\biggr)
\end{eqnarray}
Where $B_0 \textrm{ and } C_0$ are the Passarino-Veltman functions \cite{Passarino1,Passarino2}.\\
Following the on-shell renormalization scheme, there are six 
renormalization constants to compute : 
$\delta m_h^2, \delta m_W^2, \delta m_Z^2, \delta t, \delta Z_e, \delta Z_h $
 and  $\delta s_W$. $\delta Z_{AA} \textrm{ and } \delta Z_{ZA}$ 
are the 
field renormalization constant for the photon and $Z-\gamma$ 
mixing respectively and are given by :
\begin{eqnarray}
\delta Z_{AA} =\frac{\alpha}{\pi}\frac{\partial}{\partial p^2}
 B_{00}(p^2,m_{H^\pm}^{2},m_{H^\pm}^{2})\Big|_{p^2=0},\quad 
 \delta Z_{ZA} =\frac{\alpha (1-2 s_{W}^2)}{2m_{Z}^2c_{W}\pi}\biggl(
-A_{0}(m_{H^\pm}^{2})+2 B_{00}(0,m_{H^\pm}^{2},m_{H^\pm}^{2})\biggr)
\end{eqnarray}
The electric charge renormalization constant as well as the renormalization constants for the $W$ and $Z$ masses are given by:
\begin{eqnarray}
 \delta Z_e &=& -\frac{1}{2}\biggl(\delta Z_{AA}-\frac{s_{W}}{c_{W}} 
\delta Z_{ZA}\biggr),\\ 
  \delta m_{W}^2 &=& \frac{\alpha}{16 \pi s_{W}^2}\biggl(2A_{0}(m_{H^\pm}^{2})+A_{0}(m_{A^0}^{2})+A_{0}(m_{H^0}^{2})
 +4B_{00}(m_{W}^{2},m_{H^\pm}^{2},m_{H^0}^{2})\biggr) 
\end{eqnarray}
\begin{eqnarray}
 \delta m_{Z}^2 &=&- \frac{\alpha}{16 \pi s_{W}^2 c_{W}^2}\biggl(-A_{0}(m_{A^0}^{2})-A_{0}(m_{H^0}^{2})-
2(1-2s_{W}^2)^2A_{0}(m_{H^\pm}^{2}) 
 +4B_{00}(m_{Z}^{2},m_{H^0}^{2},m_{A^0}^{2}) \nonumber \\&& +4(1-2s_{W}^2)^2B_{00}(m_{Z}^{2},m_{H^\pm}^{2},m_{H^\pm}^{2} )\biggr)
 \end{eqnarray}
The counter term $\delta s_W$, corresponding to the Weinberg mixing 
angle, is determined from the tree level relation $s_W^2 = 1 - m_W^2/m_Z^2$.
$\delta s_W$ is given by :
\begin{eqnarray}
 \delta s_{W}=\frac{-c_{W}^2}{2 s_{W}}\biggl(\frac{\delta m_{W}^2}{m_{W}^2}-
 \frac{\delta m_{Z}^2}{m_{Z}^2}\biggr)
\end{eqnarray}
 The counter-terms for the Higgs mass, Higgs field and the tadpole are given by :
 \begin{eqnarray}
 \delta m_{h}^2 &=&\frac{1}{32 \pi}\biggl(2 \lambda_3 A_{0}(m_{H^\pm}^{2})+(\lambda_3+\lambda_4+\lambda_5)
A_{0}(m_{H^0}^2)+ (\lambda_3+\lambda_4-\lambda_5)A_{0}(m_{A^0}^{2})\biggr) \nonumber \\
&+&\frac{m_{W}^2 s_{W}^2}{32 \alpha \pi^3}\biggl(2 \lambda_3 ^2 B_{0}(m_{h}^{2},m_{H^\pm}^{2},m_{H^\pm}^{2})
+(\lambda_3+\lambda_4+\lambda_5)^2 B_{0}(m_{h}^{2},m_{H^0}^{2},m_{H^0}^{2})
\nonumber \\
&+&(\lambda_3+\lambda_4-\lambda_5)^2
B_{0}(m_{h}^{2},m_{A^0}^{2},m_{A^0}^{2})\biggr) \\ 
 \delta Z_h &=&\frac{m_{W}^2 s_{W}^2}{32 \alpha \pi^3}\biggl(2 \lambda_3 ^2 \frac{\partial}{\partial p^2}
 B_{0}(p^2,m_{H^\pm}^{2},m_{H^\pm}^{2})+(\lambda_3+\lambda_4+\lambda_5)^2\frac{\partial}{\partial p^2}
 B_{0}(p^2,m_{H^0}^{2},m_{H^0}^{2})\nonumber \\ &+&(\lambda_3+\lambda_4-\lambda_5)^2
 \frac{\partial}{\partial p^2}B_{0}(p^2,m_{A^0}^{2},m_{A^0}^{2})\biggr)\Big|_{p^{2}=m_h^{2}}\\
 \delta t &=& -\frac{m_{W} s_{W}}{16 e \pi^2}\frac{1}{32 \pi}\biggr(2 \lambda_3 A_{0}(m_{H^\pm}^{2})+(\lambda_3+\lambda_4+\lambda_5)
A_{0}(m_{H^0}^2)+ (\lambda_3+\lambda_4-\lambda_5)A_{0}(m_{A^0}^{2})\biggr)
\end{eqnarray}
The expression for the triple Higgs coupling counter term is given 
in the third section eq.~(\ref{cthhh}). 
The renormalized triple Higgs coupling is given by :
\begin{eqnarray*}
 \widetilde{\Gamma}_{hhh}^{loop}=\Gamma_{hhh}^{loop}+ \delta \Gamma_{hhh}^{loop}
\end{eqnarray*}
We have checked that the renormalized amplitude is UV-finite and 
furthermore independent of the renormalization scale $\mu$. We derive an approximate formula for  $\widetilde{\Gamma}_{hhh}^{loop}$
In the limit $m_{\phi} = m_H = m_{A^0} = m_{H^\pm}$ and find a good agreement
with full expression.
\begin{eqnarray*}
 \widetilde{\Gamma}_{hhh}^{loop}(q^2, m^2_h, m^2_{\phi}) &\approx&
\frac{1}{8\alpha e \pi^3 q^2}\Bigg[ 3\alpha\pi q^2 m_W
   s_W\left(2\lambda^2_3 + 2 \lambda_3\lambda_4 + \lambda^2_4 +
   \lambda^{2}_5 \right)\left(x_1 \log(\frac{-1+x_1}{x_1}) + x_2
   \log(\frac{-1+x_2}{x_2}) \right)\\ \nonumber &+& 
\left(2\lambda^3_3 + 3\lambda^2_3 \lambda_4 + 3 \lambda_3\lambda^2_4 + \lambda^3_4 + 3(\lambda_3 +
   \lambda_4)\lambda^{2}_5 \right) m^2_W s^2_W \log^2(-\frac{q^2}{m^2_{\phi}})
\Bigg]\\ \nonumber &+& 
\frac{x_1 x_2}{64\alpha m^3_W m^2_Z \pi^3 s^5_W (x_1 - x_2)}\Bigg((2\lambda^2_3 +
2\lambda_3 \lambda_4 + \lambda^2_4 + \lambda^{2}_5 )(15em^4_W m^2_Z)
\\\nonumber &+& 4e\alpha^2 m^2_h \pi^2 \Big(m^2_Z (m^2_h - 2 m^2_W +
m^2_Z)(-1+2s^2_W)+m^2_{\phi}(-2m^2_Z + m^2_W (1+(c^2_W - s^2_W)^2) \\ \nonumber&+& 4m^2_Z
s^2_W)\Big)\Bigg)\\ \nonumber &\times &\left((-1+x_2)\log(\frac{-1+x_2}{x_2})
+ (1-x_1)\log(\frac{-1+x_1}{x_1})\right)\\ \nonumber &+&
\frac{x_1 x_2}{64m^3_W\pi s^5_W (x_1 - x_2)}\Bigg( \alpha e m^2_h (-2m^2_Z + m^2_W
(3+(c^2_W - s^2_W)^2 - 4s^2_W) \\ \nonumber &+& 4m^2_Z s^2_W + m^2_h
(-2+4s^2_W))\Bigg)\\ \nonumber &\times & \left(x_1 \log(\frac{-1+x_2}{x_2}) - x_2
\log(\frac{-1+x_1}{x_1})\right)\\ \nonumber &+&
\frac{\alpha^2 m^2_h m^2_Z\pi^2}{16em^3_W m^2_Z\pi^2 s^5_W}\Bigg[-2m^2_Z +
  m^2_W(3+(c^2_W - s^2_W)^2 - 4s^2_W) + 4m^2_Z s^2_W + m^2_h (-2 +
  4s^2_W)\Bigg]\\ \nonumber &\times&\left(x^3_1 \log(\frac{-1+x_1}{x_1}) - x^3_2 \log(\frac{-1+x_2}{x_2})\right)
\end{eqnarray*}
where $x_{1,2}$ are given by:
\begin{eqnarray}
x_{1,2} = \frac{1\mp\sqrt{1-4m^2_{\phi}/q^2}}{2}
\end{eqnarray}

\end{document}